\documentclass{egpubl}

\preprint

\CGFStandardLicense
\usepackage[T1]{fontenc}
\usepackage{dfadobe}  

\biberVersion
\BibtexOrBiblatex
\usepackage[backend=biber,bibstyle=EG,citestyle=alphabetic,maxbibnames=99,giveninits=true]{biblatex} 
\electronicVersion
\PrintedOrElectronic
\ifpdf \usepackage[pdftex]{graphicx} \pdfcompresslevel=9
\else \usepackage[dvips]{graphicx} \fi

\usepackage{egweblnk} 
\usepackage{float}
\usepackage{subcaption}
\usepackage{booktabs}
\usepackage{multirow}
\usepackage{amsmath}
\usepackage{amsfonts}
\usepackage{mathtools}
\usepackage{arydshln}
\usepackage{microtype}
\usepackage[ruled]{algorithm2e} \DontPrintSemicolon
\usepackage[x11names]{xcolor}

\usepackage[nameinlink, capitalise, noabbrev]{cleveref}
\crefname{algocf}{alg.}{algs.}
\Crefname{algocf}{Algorithm}{Algorithms}

\title{Real-Time Rendering of Dynamic Line Sets using Voxel Ray Tracing}

\author[B. Kraaijeveld, A. C. Jalba, A. Vilanova \& M. Chamberland]
{\parbox{0.9\textwidth}{\centering
    B. Kraaijeveld$^{1}$\orcid{0009-0005-5638-4984},
    A.C. Jalba$^{1}$\orcid{0000-0001-9821-0767},
    A. Vilanova$^{1}$\orcid{0000-0002-1034-737X} \&
    M. Chamberland$^{1}$\orcid{0000-0001-7064-0984}} \\
{\parbox{0.9\textwidth}{\centering $^1$Department of Mathematics and Computer Science, Eindhoven University of Technology, The Netherlands}}
}

\begin{document}

\captionsetup{labelfont=bf,textfont=it}

% uncomment for using teaser
\teaser{
    \vspace{-2\baselineskip}
    \centering
    \begin{subfigure}[t]{0.5\linewidth}
        \centering
        \includegraphics[width=\linewidth]{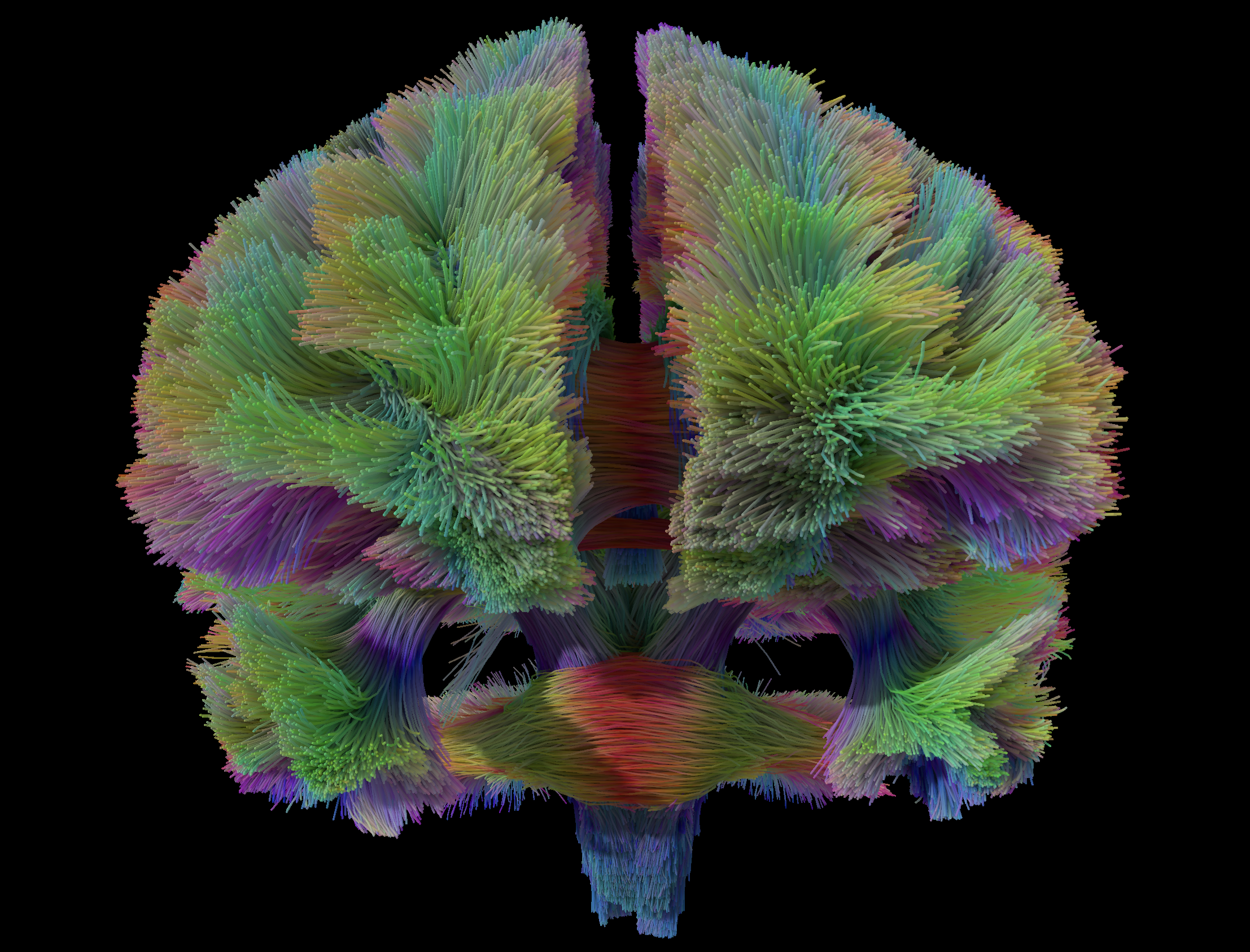}
    \end{subfigure}%
    \begin{subfigure}[t]{0.5\linewidth}
        \centering
        \includegraphics[width=\linewidth]{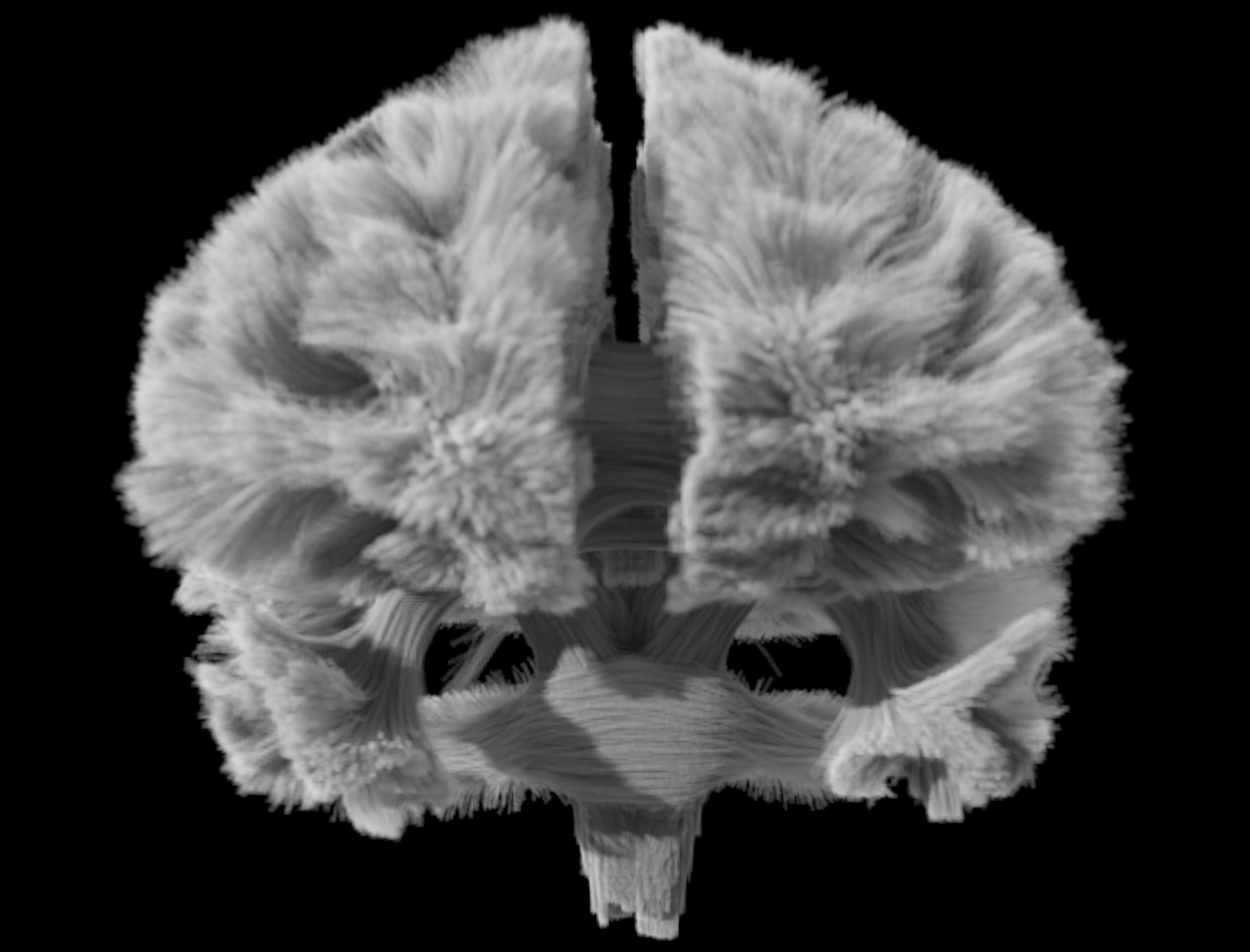}
    \end{subfigure}
    \caption{Renderings of the Bundles Large tractography line set. Left: a voxel-based ray-traced rendering with lines colored by their tangent direction as is customary in tractography. Right: a volume render of the per-voxel occupancy as produced by our voxelization method.
    Both renderings are shaded using voxel-cone-traced ambient occlusion and directional shadows.}
    \label{fig:teaser}
}

\maketitle
%-------------------------------------------------------------------------
\begin{abstract}
Real-time rendering of dynamic line sets is relevant in many visualization tasks, including unsteady flow visualization and interactive white matter reconstruction from Magnetic Resonance Imaging.
High-quality global illumination and transparency are important for conveying the spatial structure of dense line sets, yet remain difficult to achieve at interactive rates.
We propose an efficient voxel-based ray-tracing framework for rendering large dynamic line sets with ambient occlusion and ground-truth transparency.
We introduce a voxelization algorithm that supports efficient on-the-fly construction of acceleration structures for both voxel cone tracing and ray tracing.
To further reduce per-frame preprocessing cost, we propose a voxel-based culling method that restricts acceleration structure construction to camera-visible voxels.
Together, these contributions enable real-time rendering of large-scale dynamic line sets with high quality and physically accurate transparency.
We demonstrate that our method outperforms the state of the art in quality and performance when rendering (semi-)opaque dynamic line sets.


\newcommand\HH{% horrible hack
  \global\let\savedtextbullet\textbullet
  \gdef\textbullet{%
    \par\noindent\savedtextbullet\global\let\textbullet\savedtextbullet
  }%
}



\end{abstract}  
%-------------------------------------------------------------------------

\section{Introduction}

% Why Visualize Dynamic Line Sets %
Real-time rendering of 3D line sets has many applications in scientific visualization and in the game industry.
While some line sets are inherently static, others are dynamic: they vary over time as a result of a simulation, animation, or user interaction.
Visualizing dynamic line sets is relevant when dealing with time-varying flow data \cite{mcloughlin2010over}, particle trajectories, hair and fur simulation \cite{ward2007survey}, and when supporting user interactions such as segmenting diffusion MRI whole-brain tractograms into white matter bundles \cite{anwander2009white, schultz2008virtual} or real-time tractography \cite{chamberland2014real}.
Global illumination effects, such as ambient occlusion and soft shadows, facilitate depth and shape perception \cite{diaz2017experimental}, while transparency helps reveal otherwise occluded structures \cite{kern2020}.

% Transparency Challenges %
While global illumination effects can be approximated efficiently and accurately using \textit{voxel cone tracing} \cite{crassin2011, staib2015, hermosilla2016high, kanzler2018, gross2020}, transparency remains a key challenge when rendering large, dense line sets in real-time.
Accurate transparency requires compositing all geometry in visibility order, which is computationally expensive for dense line sets.
Two key strategies are commonly used to resolve transparency:
\textit{Object-order} rendering techniques process geometry element by element and rely on various techniques to capture and sort all geometry per pixel.
Gro{\ss} and Gumhold \cite{gross2020} present an object-order method that renders transparent line sets by sorting all geometry in visibility order every frame.
However, due to the sorting overhead, their method is less efficient for (semi-)opaque datasets.
\textit{Image-order} rendering methods precompute various acceleration structures and resolve transparency by tracing rays from the camera's perspective.
Kanzler et al. \cite{kanzler2018} present a voxel-based quantized line representation that allows for efficient ray-traced transparency, but relies on a precomputed acceleration structure.
While ray tracing more naturally handles transparency, there is little to no opportunity to precompute these acceleration structures in the context of dynamic line sets.

% Contribution %
In this work, we present an efficient rendering pipeline for high-quality ray tracing of dynamic line sets that computes all acceleration structures on the fly.
We achieve real-time performance by leveraging a fast voxelization scheme, which we use to accelerate the remainder of the rendering pipeline.
Based on this voxelization, we introduce a voxel-based culling method to construct a ray-tracing acceleration structure on the fly only for voxels visible to the camera.
Finally, we use these acceleration structures to render large line sets with voxel-based shading and ground-truth transparency.
While we present a pipeline to accelerate rendering of dynamic line sets, our pipeline could be extended to render almost any geometric primitive.
To summarize, our main contribution is a high-performance voxel-based ray tracing pipeline, which is made possible by:

\begin{itemize}
    \item A GPU-based method for efficient and high-quality conservative voxelization of capsule primitives.
    \item A voxel-based culling approach that only constructs a ray-tracing acceleration structure for voxels visible to the camera, saving memory and increasing performance.
    \item A voxel-based ray-tracing technique that renders ground-truth transparency with high performance for (semi-)opaque line sets.
\end{itemize}

%-------------------------------------------------------------------------
\section{Related work}

Our work is closely related to prior research line set rendering, voxelization, and transparency techniques.

\subsection{Rendering of Line Sets}

% Object-Order Techniques
In object-order rendering, accurate transparency is achieved by compositing geometry in visibility order using order-dependent and order-independent techniques.
Depth peeling \cite{everitt2001interactive} is an order-independent technique that resolves transparency by rasterizing all geometry as many times as the maximum depth complexity.
However, since a single rasterization pass is already computationally expensive for dense line sets, depth peeling has poor performance.
Staib et al. \cite{staib2015} achieve transparency in a single rasterization pass by sorting all geometry in visibility order, which is sufficient when rendering small spherical particles.
However, when rendering line segments, transparency has to be resolved for each individual pixel.
Gro{\ss} and Gumhold \cite{gross2020} combine order-dependent and order-independent transparency techniques to render accurate transparency for line segments in a single rasterization pass.
While object-order methods can achieve accurate transparency and support dynamic data, sorting and other transparency techniques incur a substantial performance overhead compared to image-order methods.

% Image-Order Techniques
Image-order methods are a more natural fit for rendering transparency, but often rely on precomputed acceleration structures, which limits their use to static datasets.
Schussman and Ma \cite{schussman2004anisotropic} precompute a per-voxel spherical harmonics representation that allows for efficient rendering of large line sets.
Kanzler et al. \cite{kanzler2018} precompute a voxel-based line segment quantization and a level-of-detail hierarchy, enabling efficient rendering of large transparent line sets at the cost of accuracy.
McGraw \cite{mcgraw2020} precomputes a sparse voxel octree to accelerate ray marching of high-quality interpolated line sets with ray-marched shadows.
Wald et al. \cite{wald2020} repurposed GPU ray-tracing hardware to accelerate the ray tracing of line sets based on a bounding volume hierarchy of oriented bounding boxes.
Our method is most similar to that of Kanzler et al. \cite{kanzler2018} but supports dynamic datasets by constructing all acceleration structures on the fly.
Furthermore, we improve the rendering quality and support ground truth transparency, while being as efficient in rendering dynamic (semi-)opaque line sets.

% Voxel Cone Tracing
Global illumination for line sets is tackled using, e.g., ray tracing \cite{wald2020}, sparse voxel octrees \cite{mcgraw2020}, or screen-space techniques \cite{eichelbaum2012}.
Voxel cone tracing \cite{crassin2011} is particularly well suited for rendering soft shadows and ambient occlusion for dense line sets \cite{kanzler2018, gross2020} and molecular datasets \cite{staib2015, hermosilla2016high}.
Voxel cone tracing provides great performance, while delivering high-quality low-frequency shading, which improves visual clarity over high-frequency shading techniques \cite{kanzler2018}.
In our work, we repurpose the voxel cone tracing acceleration structure to accelerate rendering by employing voxel-based primitive culling, texture-space shading, and empty-space skipping.

\subsection{Voxelization of Line Sets}

% Occupancy Pyramid and Voxelization
Voxel cone tracing relies on an acceleration structure called an \textit{occupancy pyramid} \cite{hermosilla2016high}: a 3D texture hierarchy where each voxel encodes the fraction of its volume occupied by geometry.
In prior work, the occupancy pyramid is often constructed using voxelization techniques \cite{staib2015, hermosilla2016high, kanzler2018, gross2020}.

% Early Voxelization of Line Segments %
Voxelization of line segments has been a longstanding topic in computer graphics.
An early influential algorithm is Bresenham's line drawing algorithm \cite{bresenham1965}, which may be extended to 3D for voxelization.
While Bresenham's algorithm produces a single-pixel-width line, it is not conservative, i.e., not all pixels/voxels intersected by the line are recognized.
Xiaolin Wu proposed a method to rasterize anti-aliased lines \cite{wu1991efficient}, removing the staircasing effect at the cost of being more expensive than Bresenham.
Amanatides \& Woo's Digital Differential Analyzer (DDA) algorithm \cite{amanatides1987} provides a conservative voxelization for infinitely thin line segments and is widely used as a ray-voxel traversal algorithm.
However, none of these algorithms guarantee a conservative voxelization of a line segment with a radius, limiting their applicability for constructing ray-tracing acceleration structures.

% Voxelization of Line Segments in the context of Voxel Cone Tracing %
Voxelization of line segments can be used to generate an occupancy pyramid to enable real-time voxel cone tracing, as is shown by Gro{\ss} and Gumhold \cite{gross2020} and Kanzler et al. \cite{kanzler2018}, who respectively use rounded cone and cylinder primitives during rendering, but infinitely thin lines during voxelization.
Both demonstrate how non-conservative voxelization of line sets can generate high-quality occupancy pyramids.
Kanzler et al. \cite{kanzler2018}, however, show non-conservative voxelization leads to missing-segment artifacts when used as the basis for ray tracing.
While they address these artifacts by including all voxel neighbors in ray-tracing intersection tests, this approach adds a significant performance overhead.
We demonstrate how our conservative capsule voxelization addresses these artifacts, while maintaining high ray-tracing performance.

% Voxelization In Molecular Visualization
Molecular rendering methods generate an occupancy pyramid every frame to enable real-time voxel cone tracing for dynamic data \cite{staib2015, hermosilla2016high}.
Staib et al. \cite{staib2015} voxelize spherical particles by rendering planar slices to a 3D texture, using multi-sampling to create a conservative, anti-aliased voxelization.
Hermosilla et al. \cite{hermosilla2016high} voxelize molecules that consist of atoms and bonds, modeled by sphere and cylinder primitives, respectively.
They estimate occupancy by evaluating the primitive's signed distance function at the center of each voxel within the primitive's \textit{Axis-Aligned Bounding Box} (AABB).
Like Hermosilla et al. \cite{hermosilla2016high} and Hsieh et al. \cite{hsieh2010}, we use signed distance functions to estimate occupancy.
However, since our line sets contain line segments that may be long, thin, and not axis-aligned, it can be expensive to evaluate all voxels within the primitive's AABB.
To improve performance for conservative voxelization of line primitives, we present a voxel traversal algorithm that scales better with line segment length.

\subsection{Order-Independent Transparency}

A-Buffer techniques \cite{carpenter1984buffer} improve performance over depth peeling by capturing multiple fragments per pixel in a single rasterization pass.
A key challenge in A-Buffer techniques is the unknown number of fragments per pixel, which results in unbounded memory cost and indeterminate fragment write locations.
Per-Pixel Linked Lists \cite{yang2010real} capture all fragments in a single rendering pass but introduce significant memory overhead due to storing a node pointer with every fragment.
Fragment Pages reduce this overhead by providing GPU support for unrolled linked lists \cite{Crassin}.
The Linearized Layered Fragment Buffer \cite{knowles2012efficient} instead captures all fragments in two rendering passes.
The first pass computes per-pixel fragment counts, which are converted into per-pixel memory offsets using a parallel prefix sum \cite{harris2012optimizing}.
A secondary rendering pass then captures all fragments into contiguous memory.
Because these A-Buffer methods capture all fragments per pixel, they introduce significant memory and bandwidth costs for scenes with high depth complexity.
Furthermore, while these methods capture all fragments, they still need to be sorted to produce correct transparency.
Because generally more fragments are generated than there are primitives, sorting fragments is more expensive than sorting primitives.
In our work, we extend the Linearized Layered Fragment Buffer \cite{knowles2012efficient} for the voxel domain with a two-pass voxelization scheme to capture per-voxel fragment lists.
By only capturing per-voxel fragment lists for the voxels visible to the camera, our method saves memory and improves performance when rendering dynamic line sets.

%-------------------------------------------------------------------------
\section{Voxel-Based Ray Tracing of Dynamic Line Sets} \label{method:overview}

\begin{figure*}
    \centering
    \includegraphics[width=\linewidth]{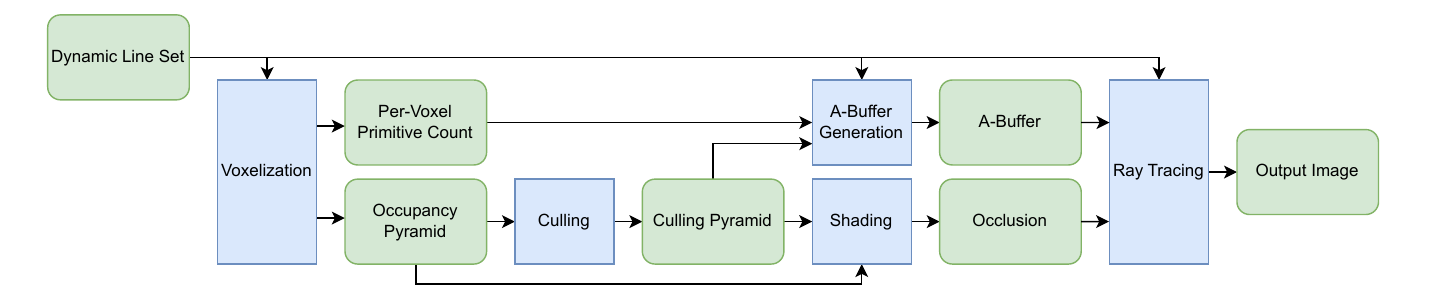}
    \caption{Schematic overview of our rendering pipeline.
    The stages of our rendering pipeline are represented by the blue blocks and are described in \cref{method:voxelization,method:culling,method:abuffer,method:shading,method:rendering}.
    Green blocks represent data structures, while arrows show the inputs and outputs of each pipeline stage.}
    \label{fig:method:pipeline-overview}
    \vspace{-\baselineskip}
\end{figure*}

In this section, we describe our rendering pipeline for real-time visualization of dynamic line sets with ambient occlusion and transparency.
An overview of the pipeline is provided in \cref{fig:method:pipeline-overview}.

We voxelize all line segments every frame to compute the \textit{occupancy pyramid} and \textit{per-voxel primitive counts}, see \cref{method:voxelization}.
Based on this voxelization, we use voxel-based culling to determine the camera-visible voxels, see \cref{method:culling}.
Then, we compute \textit{per-voxel fragment lists} for just the camera-visible voxels, described in \cref{method:abuffer}.
We compute shading in voxel space using voxel cone tracing, see \cref{method:shading}.
Finally, we ray trace line sets with ground truth transparency and ambient occlusion, see \cref{method:rendering}.

\subsection{Voxelization} \label{method:voxelization}

To support real-time voxel cone tracing for dynamic line sets, we voxelize the entire line set every frame to generate the \textit{occupancy pyramid}.
To enable ray tracing, we use conservative voxelization to capture all line–voxel intersections.
Each intersection contributes to the \textit{per-voxel primitive count} and the estimated voxel \textit{occupancy}.

\begin{figure}
    \centering
    \includegraphics[width=\linewidth]{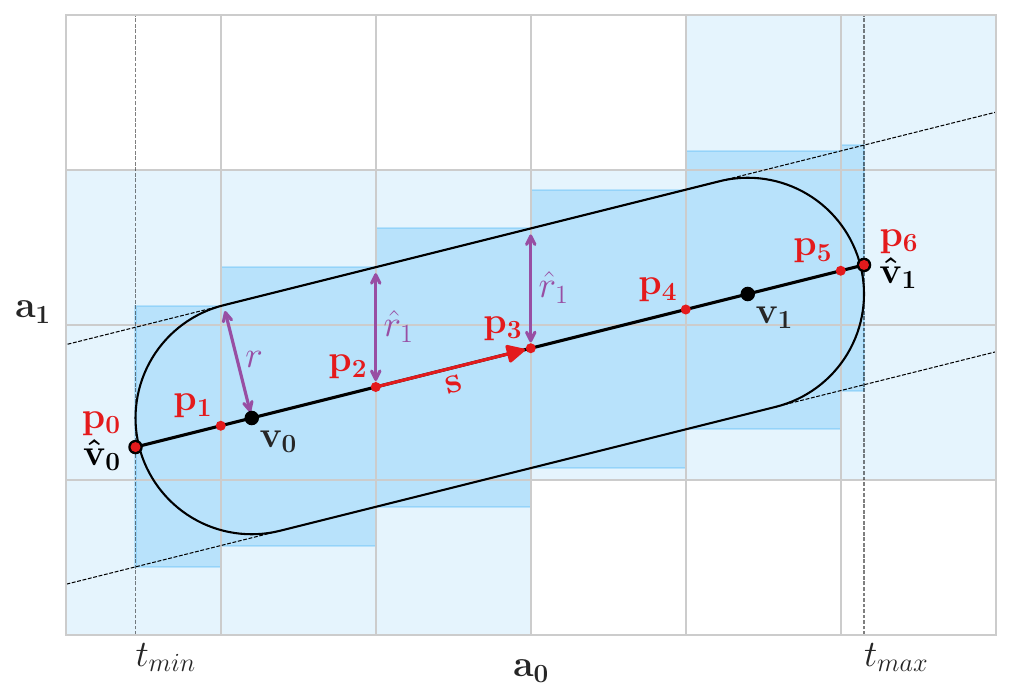}
    \caption{
        2D Illustration of our voxel traversal scheme.
        Our algorithm traverses axis \(a_0\), visiting points \(\mathbf{p_0}\)--\(\mathbf{p_6}\). For each consecutive pair of points, a bounding box (dark blue) is computed along the minor axes \(a_1, a_2\) using the projected radii \(\hat{r}_1, \hat{r}_2\). All intersected voxels (light blue) are voxelized. See the text for details.}
    \label{fig:method:voxelization}
\end{figure}

During voxelization, we represent each line segment as a Quilez-style capsule primitive \cite{quilezDistFunctions}, defined by vertices \(\mathbf{v_0}\) and \(\mathbf{v_1}\) and radius \(r\).
Capsule primitives connect adjacent line segments without leaving gaps but constrain each polyline to have a uniform radius.
We traverse the axis corresponding to the largest-magnitude component---\(x\), \(y\) or \(z\)---of the line delta \(\mathbf{d} = \mathbf{v_1} - \mathbf{v_0}\).
At each step along the major axis, we visit the voxels along the remaining two (minor) axes.
Unlike a digital differential analyzer (DDA), our method ensures conservative voxelization for capsules, while visiting fewer voxels than axis-aligned bounding box (AABB) voxelization.
\cref{fig:method:voxelization} shows an illustration of our voxel traversal algorithm.

We sort the components of the line delta $\mathbf{d}$ by absolute magnitude to obtain the axis index vector $\mathbf{a} = (a_0, a_1, a_2)$, where $a_0$ denotes the major axis and $a_1, a_2$ the minor axes.
To simplify traversal, we swap vertices \(\mathbf{v_0}\) and \(\mathbf{v_1}\) if the major axis has a negative direction.
By normalizing \(\mathbf{d}\) relative to its largest vector component \(d_{a_0}\), we obtain a step vector \(\mathbf{s} = \mathbf{d}/d_{a_0}\) as the vector from one major axis voxel boundary to the next.
To account for the capsule end caps, we extend the line segment in both directions with \(\mathbf{s} \cdot r\) to obtain the extended vertices \(\mathbf{\hat{v}_0}\) and \(\mathbf{\hat{v}_1}\).
Finally, we compute the projected capsule radii $\hat{r}_1$ and $\hat{r}_2$ along the minor axes $a_1$ and $a_2$ using:

\vspace*{-\baselineskip}
\begin{align}
    \hat{r}_i &= \dfrac{r}{\sqrt{1 - \left(d_{a_i}/\left\|\mathbf{d}\right\|\right)^2}} \quad\text{where } i \in 1,2.
\end{align}

We traverse along the major axis \(a_0\), from extended vertex \(\mathbf{\hat{v}_0}\) to \(\mathbf{\hat{v}_1}\), and visit every voxel boundary in between.
For each consecutive pair of voxel intersection points along the major axis \(a_0\), defined as \(\mathbf{p_n}\) and \(\mathbf{p_{n+1}}\), we compute a 2D bounding box in the minor axes \(a_1, a_2\) using radii \(\hat{r}_1, \hat{r}_2\).
We visit all voxels that overlap this bounding box.
The voxel traversal pseudocode is provided in \cref{algorithm:voxelization}.

\begin{algorithm}
\caption{Conservative Capsule Voxelization}
\label{algorithm:voxelization}
\SetKwFunction{Visit}{Visit}

$t_{min} \gets \hat{v}_{0,a_0}$ \tcp{major axis component of \(\hat{v}_{0}\)}
$t_{max} \gets \hat{v}_{1,a_0}$ \tcp{major axis component of \(\hat{v}_{1}\)}
$t_0 \gets t_{min}$\;
$\mathbf{p_0} \gets \mathbf{\hat{v}_0}$\;
\;
\While{$t_0 < t_{max}$} {
    \tcp{Compute next intersection point}
    $t_1 \gets \min(t_{max}, \text{floor}(t_0 + 1))$\;
    $\mathbf{p_1} \gets \mathbf{\hat{v}_0} + \mathbf{s} (t_1 - t_{min})$\;
    \;
    \tcp{Define 2D box to voxelize}
    $j_{min} \gets \text{int}(\min(p_{0,a_1}, p_{1,a_1}) - \hat{r}_1)$\;
    $j_{max} \gets \text{int}(\max(p_{0,a_1}, p_{1,a_1}) + \hat{r}_1)$\;
    $k_{min} \gets \text{int}(\min(p_{0,a_2}, p_{1,a_2}) - \hat{r}_2)$\;
    $k_{max} \gets \text{int}(\max(p_{0,a_2}, p_{1,a_2}) + \hat{r}_2)$\;
    \;
    \tcp{Visit all voxels within 2D box}
    \For{$j=j_{min}$ \KwTo $j_{max}$, $k=k_{min}$ \KwTo $k_{max}$} {
        $\mathbf{C}_{a_0} = \text{int}(t_0)$,
        $\mathbf{C}_{a_1} = j$,
        $\mathbf{C}_{a_2} = k$\;
        \Visit{$\mathbf{C}$}\; 
    }
    \;
    \tcp{Move to next intersection point}
    $t_0 \gets t_1$\;
    $\mathbf{p_0} \gets \mathbf{p_1}$\;
}
\end{algorithm}

% Theoretical Efficiency
While AABB voxelization exhibits linear, quadratic, or cubic time complexity depending on line orientation,
like DDA, our capsule voxelization scales linearly with line segment length, regardless of line orientation.
Because most line sets contain a wide range of line orientations, our method is expected to outperform AABB when voxelizing longer line segments.

\begin{figure}
    \centering
    \includegraphics[width=\linewidth]{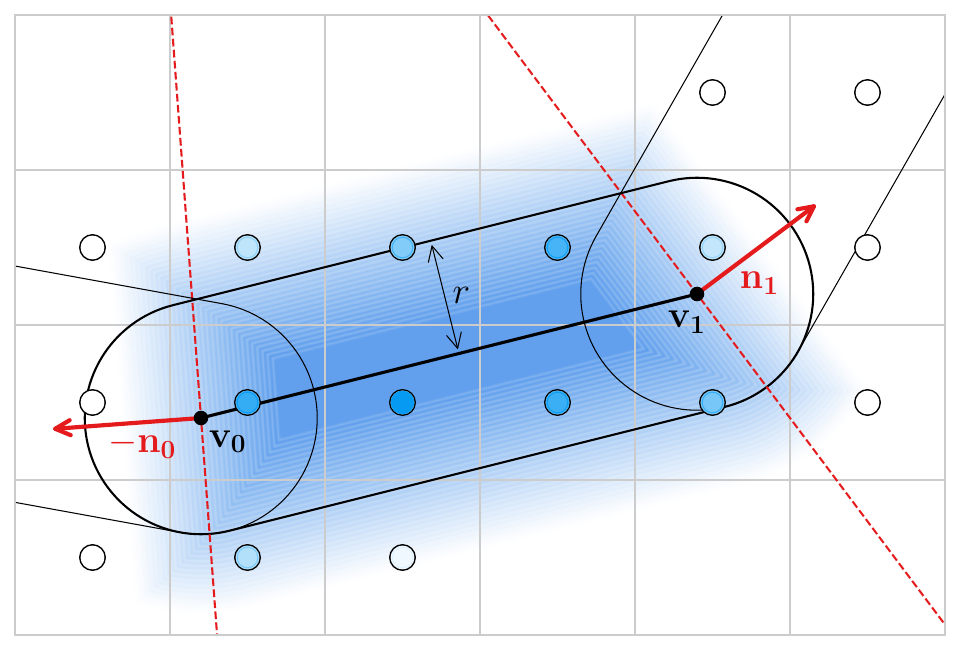}
    \caption{
        2D Illustration of our SDF-based occupancy estimation.
        To compute the occupancy for a given capsule-voxel intersection, the clipped capsule occupancy function (\cref{algorithm:capsule:occupancy}, visualized as a blue gradient) is sampled at the center of each intersecting voxel (filled white-blue circles). See text for more details.}
    \label{fig:method:occupancy}
\end{figure}

% Signed Distance Function Approximation of Occupancy %
To compute the base level of the occupancy pyramid, we iterate over all capsule-voxel intersections using our voxel traversal algorithm.
As no closed-form expression for the capsule-voxel intersection volume exists, we approximate occupancy using the primitive's signed distance function, following Hermosilla et al. \cite{hermosilla2016high} and Hsieh et al. \cite{hsieh2010}.
% Signed Distance Function of Clipped Capsule Primitives %
However, when representing line segments as capsules, the endpoints of adjacent capsules overlap.
This results in spherical artifacts at each interior vertex, which lead to an overestimation of occupancy, particularly for densely packed vertices.
To prevent overlapping geometry, we follow Gro{\ss} and Gumhold \cite{gross2020} in defining a clipping plane normal $\mathbf{n_i}$ as the tangent at $\mathbf{v_i}$ computed via central differences.
In \cref{algorithm:capsule}, we extend the Quilez capsule SDF \cite{quilezDistFunctions} to define a clipped capsule SDF.
This SDF computes the signed distance from point \(\mathbf{p}\) to a clipped capsule with vertices \(\mathbf{v_0}, \mathbf{v_1}\), clipping planes \(\mathbf{n_0}, \mathbf{n_1}\) and radius \(r\).
The signed-distance-based occupancy estimation is illustrated in \cref{fig:method:occupancy}.

\begin{algorithm}
\caption{Clipped Capsule Signed Distance Function}
\label{algorithm:capsule}
\SetKwFunction{CapsuleSDF}{ClippedCapsuleSDF}
\SetKwProg{Fn}{Function}{:}{}

\Fn{\CapsuleSDF{$\mathbf{p}, \mathbf{v_0}, \mathbf{v_1}, \mathbf{n_0}, \mathbf{n_1}, r$}}{
    $\mathbf{d} \gets \mathbf{v_1} - \mathbf{v_0}$\;
    $\mathbf{pv_0} \gets \mathbf{p} - \mathbf{v_0}$\;
    $\mathbf{pv_1} \gets \mathbf{p} - \mathbf{v_1}$\;
    $h \gets \text{clamp}\left(\left(\mathbf{pv_0} \cdot \mathbf{d}\right) / \left(\mathbf{d} \cdot \mathbf{d}\right), 0, 1\right)$\;
    \;
    $sdf_{capsule} \gets \|\mathbf{pv_0} - \mathbf{d} \cdot h\| - r$\;
    $sdf_{n_0} \gets -\mathbf{pv_0} \cdot \mathbf{n_0}$\;
    $sdf_{n_1} \gets \mathbf{pv_1} \cdot \mathbf{n_1}$\;
    \;
    \Return $\max(sdf_{capsule}, sdf_{n_0}, sdf_{n_1})$\;
}
\end{algorithm}

% Telephone Wire Anti-Aliasing %
The clipped capsule signed distance function provides a good occupancy approximation for capsules with large radii relative to voxel size.
However, aliasing is introduced when the capsule radius is smaller than half the voxel size.
To reduce aliasing, we adapt Phone Wire Anti-Aliasing (PWAA) \cite{persson2012}, which clamps the radius of thin line segments to the pixel size and fades alpha instead.
We adapt PWAA to the voxel domain by clamping the clipped capsule radius to a minimum value \(r_{min}\) and correcting for the radius difference in the occupancy estimation.
The occupancy estimation algorithm is shown in \cref{algorithm:capsule:occupancy}.

\begin{algorithm}
\caption{Clipped Capsule Occupancy Estimation}
\label{algorithm:capsule:occupancy}
\SetKwFunction{CapsuleOccupancy}{CapsuleOccupancy}
\SetKwProg{Fn}{Function}{:}{}

    \Fn{\CapsuleOccupancy{$\mathbf{p}, \mathbf{v_0}, \mathbf{v_1}, \mathbf{n_0}, \mathbf{n_1}, r, r_{min}$}}{
    $r_{clamp} \gets \max(r, r_{min})$\;
    $r_{correction} = (r / r_{clamp})^2 $\;
    \;
    \tcp{Compute Occupancy from SDF}
    $sdf \gets \text{ClippedCapsuleSDF}(\mathbf{p}, \mathbf{v_0}, \mathbf{v_1}, \mathbf{n_0}, \mathbf{n_1}, r_{clamp})$\;
    $occupancy = \text{clamp}(0.5 - sdf, 0, 1)$\;
    \;
    \Return $ occupancy \cdot r_{correction}$
}
\end{algorithm}

To construct the base level of the occupancy pyramid on the GPU, each thread voxelizes a single line segment at a time, summing per-voxel occupancy using atomic operations.
To complete the occupancy pyramid, an averaging mipmap is generated.
The base level of the occupancy pyramid is shown in \cref{fig:teaser}.
To compute the first pass of the Linearized Layered Fragment Buffer \cite{knowles2012efficient}, we compute the \textit{per-voxel primitive count} alongside the occupancy pyramid.
Since our voxelization is bandwidth bound, we pack both occupancy and primitive count into a single 32-bit integer, reserving 16 bits for each value.
Only a single 32-bit atomic addition is needed to update both values for each voxel, significantly improving performance.

\subsection{Culling} \label{method:culling}

We use the occupancy pyramid to cull voxels that are not visible to the camera.
To determine visibility, we trace a ray from the center of each occupied voxel towards the camera and store the result in the base level of a new 3D texture hierarchy called the \textit{culling pyramid}.
We then generate a mipmap hierarchy in which each parent voxel encodes whether any of its child voxels are visible to the camera.

Determining visibility based on the occupancy pyramid can result in erroneously culled voxels, particularly for dense line sets viewed at grazing angles.
These errors arise from the relatively low resolution of the occupancy pyramid, which allows rays to pass through voxels that appear occupied but are partially empty.
To create a more conservative occupancy estimation, we apply morphological erosion to the base level of the occupancy pyramid by taking the minimum occupancy of each voxel's 6-neighborhood.
Tracing rays through the eroded occupancy volume yields a more conservative culling pyramid and eliminates culling artifacts.

\begin{table*}[ht!]
\centering
\begin{tabular}{l|rrll}
\textbf{line set} & \textbf{\# polylines} & \textbf{\# segments} & \textbf{avg. segment length} & \textbf{description} \\
\hline
Bundles Small       & 24.000    & 735.080     & 6.17 
& Segmented Tractogram (Four Bundles) \\
Aneurysm            & 9.213     & 2.267.219   & 1.75
& Streamlines in anterior of Aneurysm \\
Bundles Large       & 216.000   & 4.963.145   & 5.27
& Segmented Tractogram (All Bundles) \\
Brain 200k          & 200.000   & 10.846.113  & 0.97
& Whole Brain Tractogram \\
Turbulence          & 80.000    & 17.468.339  & 1.10
& Streamlines advected in forced turbulence field \\
Brain 1M            & 1.000.000 & 54.240.953  & 0.31
& Whole Brain Tractogram \\
\end{tabular}
\caption{
    Line sets used in our experiments.
    For each line set, we provide their polyline count, segment count, average segment length relative to the voxel size at a resolution of \(128^3\), and a short description. Renderings of these line sets can be found in \cref{appendix:datasets}.}
\label{tab:results:data}
\end{table*}

\subsection{Per-Voxel Fragment Lists} \label{method:abuffer}

Next, we create a ray-tracing acceleration structure by storing all fragments, i.e. capsule-voxel intersections, in \textit{per-voxel fragment lists}, where each entry records the intersecting line segment index.

A common strategy for generating per-voxel fragment lists is to sort all fragments based on the Morton-encoded voxel coordinates.
This \textit{voxelize-sort-scan} strategy is used in the context of line rendering by, e.g., Kanzler et al. \cite{kanzler2018} and McGraw \cite{mcgraw2020}.
However, even the fastest sorting algorithm comes with a notable performance overhead.
OneSweep, a state-of-the-art parallel radix-sort algorithm, sorts \(n\) fragments using \(5n\) memory operations, given 32-bit keys \cite{adinets2022onesweep}.
While many methods rely on sorting, fragment order within a voxel is inherently undefined, since it depends on the ray that intersects the voxel.
We can therefore generate the acceleration structure using a more efficient bucket sort algorithm, where we treat each voxel as a bucket.
Instead of sorting, our method extends the Linearized Layered Fragment Buffer \cite{knowles2012efficient} to the voxel domain.
This \textit{voxelize-scan-voxelize} strategy uses two voxelization passes, i.e. \(2n\) memory operations, and one scan over all voxels to construct to construct the per-voxel fragment lists.
Because we use a two-pass voxelization scheme, we can improve performance by accelerating the secondary pass based on the culling pyramid.
This results in the \textit{voxelize-cull-scan-voxelize} strategy.

To construct the per-voxel fragment lists, we first compute a parallel prefix-sum over the per-voxel primitive count buffer for all voxels marked visible in the culling pyramid.
This prefix-sum yields a \textit{per-voxel memory offset} for all camera-visible voxels.
After culling entire line segments by testing their axis-aligned bounding box against the culling pyramid, we voxelize the remaining line segments.
For all capsule-voxel intersections, we sample the base level of the culling pyramid to check whether the fragment entry should be written, preventing erroneous writes to culled voxels.
For all remaining capsule-voxel intersections, we atomically increment the per-voxel memory offset and write the line segment index.
Since we only consider polylines that neither branch nor loop, we use primitive index \(i\) to refer to the line segment from vertex \(\mathbf{v_i}\) to vertex \(\mathbf{v_{i+1}}\), which reduces the size of the per-voxel fragment lists.

\subsection{Shading} \label{method:shading}

Even though voxel cone tracing is an efficient global illumination method, evaluating it for every ray-capsule intersection becomes impractical when rendering dense line sets with transparency.
Therefore, we compute and store global illumination effects in voxel space before the ray-tracing step.
This voxel-based shading scheme makes evaluating global illumination inexpensive during ray tracing, requiring only a single texture sample per ray–capsule intersection.

We compute ambient occlusion by sampling the occlusion pyramid using voxel-cone-traced cones whose directions are equally distributed around the sphere.
Similarly, we compute directional shadows by tracing a ray from the voxel center toward the directional light source.
As an optimization, we only compute shading for voxels marked visible in the culling pyramid.

\begin{table*}
\centering
\begin{tabular}{l|c:cc|c:cc|c:cc}
\multicolumn{1}{r|}{\textbf{Resolution}} & 
\multicolumn{3}{c|}{$\mathbf{128^3}$} & 
\multicolumn{3}{c|}{$\mathbf{256^3}$} & 
\multicolumn{3}{c }{$\mathbf{512^3}$} \\
\textbf{Dataset}
 & DDA & capsule & AABB & 
   DDA & capsule & AABB & 
   DDA & capsule & AABB \\
\hline
Bundles Small       & 3.38 & \textbf{3.61}  & 4.72
                    & 4.70 & \textbf{6.20}  & 13.6
                    & 28.8 & \textbf{31.6}  & 72.4 \\
Aneurysm            & 1.29 & \textbf{2.70}  & 2.81
                    & 5.22 & 5.24           & \textbf{5.22}
                    & 29.2 & \textbf{30.8}  & 31.9 \\
Bundles Large       & 3.32 & \textbf{6.07}  & 10.7
                    & 6.14 & \textbf{12.3}  & 36.5
                    & 44.0 & \textbf{51.5}  & 179  \\
Brain 200k          & 2.70 & 5.19           & \textbf{3.92}
                    & 8.97 & 10.2           & 10.2
                    & 40.8 & \textbf{47.3}  & 48.8 \\
Turbulence          & 6.53 & 8.15           & \textbf{7.62}
                    & 22.8 & \textbf{28.0}  & 29.2
                    & 64.9 & \textbf{78.8}  & 85.2 \\
Brain 1M            & 9.28 & 20.0           & \textbf{16.7}
                    & 29.1 & \textbf{34.9}  & 35.6
                    & 114  & \textbf{146}   & 153  \\
\end{tabular}
\caption{
    Voxelization performance, measured in milliseconds, between our capsule traversal and AABB traversal, with DDA as a reference.}
\label{tab:results:voxelization:performance}
\end{table*}

\subsection{Ray-Tracing} \label{method:rendering}

Using the per-voxel fragment lists, we render all lines front to back using voxel-based ray tracing.
We trace a single ray for every pixel and use octree traversal to efficiently identify the first occupied voxel of the culling pyramid.
We then traverse the voxel grid using DDA and query each voxel's fragment list for intersecting line segments.
We use a ray-capsule intersection test \cite{quilezIntersectors}, combined with clipping planes, to compute the intersection point with the clipped capsule.
To guarantee correct hit ordering, we only consider intersection points within the current voxel.
Thanks to conservative voxelization, every ray passes through a voxel containing both the primitive index and intersection point of each intersecting primitive.

When rendering opaque geometry, only the closest intersection point within a voxel needs to be determined, after which we employ early ray termination and proceed to shading, making ray tracing especially effective for opaque geometry.
When rendering transparent geometry, we need to capture all ray-capsule hits within a voxel in visibility order.
Storing all hits in private GPU memory, however, increases register pressure and reduces performance.
We therefore encode each hit in a single 32-bit integer by storing the per-voxel hit depth in the most significant 16 bits and the per-voxel fragment index in the least significant 16 bits.
We obtain the closest \(k\) hits by insertion-sorting the depth-index keys in private memory as fragments arrive.
Only then do we evaluate shading and blend the fragments to the screen.
When a voxel contains more than \(k\) hits, we repeat the procedure, considering only fragments deeper than the \(k\)th entry from the last iteration.
Using this sorting strategy, we achieve ground-truth transparency.
\cref{fig:teaser} shows an example of a semi-transparent rendering using this technique.

%-------------------------------------------------------------------------
\section{Results \& Discussion}

% Hardware Setup and Experimental Settings
We implemented and evaluated our methods using Rust and wgpu on an Apple M3 MacBook with an 18-core GPU and 36 GB of RAM.
We have made the source code and a web demo available on \url{https://github.com/as-the-crow-flies/vibrant-tractography}.

We tested our method on several line sets, which are summarized in \cref{tab:results:data}.
Our tractography line sets are based on diffusion MRI data from the Human Connectome Project \cite{van2013wu}.
The \textit{Brain 200k} and \textit{Brain 1M} line sets were obtained using MRtrix \cite{tournier2019mrtrix3}.
The \textit{Bundles Small} and \textit{Bundles Large} line sets were obtained using TractSeg \cite{wasserthal2018tractseg}.
The \textit{Aneurysm} and \textit{Turbulence} flow line sets are part of the public dataset from Kern et al. \cite{Kern_2020}.

% Reference to upcoming sections
In \cref{results:voxelization}, we evaluate the quality and performance of our conservative capsule voxelization.
In \cref{results:abuffer}, we evaluate the performance of and the effect of culling on constructing the per-voxel fragment lists.
In \cref{results:rendering}, we evaluate and compare our overall dynamic rendering approach with the methods of Kanzler et al. \cite{kanzler2018} and Gro{\ss} and Gumhold \cite{gross2020}.

\subsection{Voxelization} \label{results:voxelization}

% Performance Analysis
We first evaluate the performance of our \textit{capsule} voxelization method against two commonly used methods:
\begin{itemize}
    \item \textbf{DDA}: the Digital Differential Analyzer algorithm by Amanatides and Woo \cite{amanatides1987}, which generates a conservative voxelization for infinitely thin lines, but is not conservative for capsule primitives.
    Occupancy is estimated as in Gro{\ss} and Gumhold \cite{gross2020}.
    \item \textbf{AABB}: Axis-Aligned Bounding Box voxelization following Hermosilla et al. \cite{hermosilla2016high}, which visits all voxels within the primitive's AABB, sampling the primitive's signed distance function (SDF) to estimate occupancy.
\end{itemize}

To compare the performance of these three voxelization strategies, we measured the voxelization time for the six line sets at various voxel-grid resolutions.
The results are shown in \cref{tab:results:voxelization:performance}.
For all tests, we used a line segment radius of \(0.2\) times the voxel size.
Since \textit{capsule} voxelization is conservative and therefore visits more voxels, it is always slower than \textit{DDA} voxelization.
For very short line segments, \textit{capsule} voxelization shows no performance advantage over \textit{AABB} voxelization and even shows reduced performance at lower voxel resolutions.
However, for the \textit{Bundles Small \& Large} line sets, which have longer line segments, a significant performance advantage is observed, especially at higher grid resolutions.

\begin{figure}
    \centering
    \includegraphics[width=0.9\linewidth]{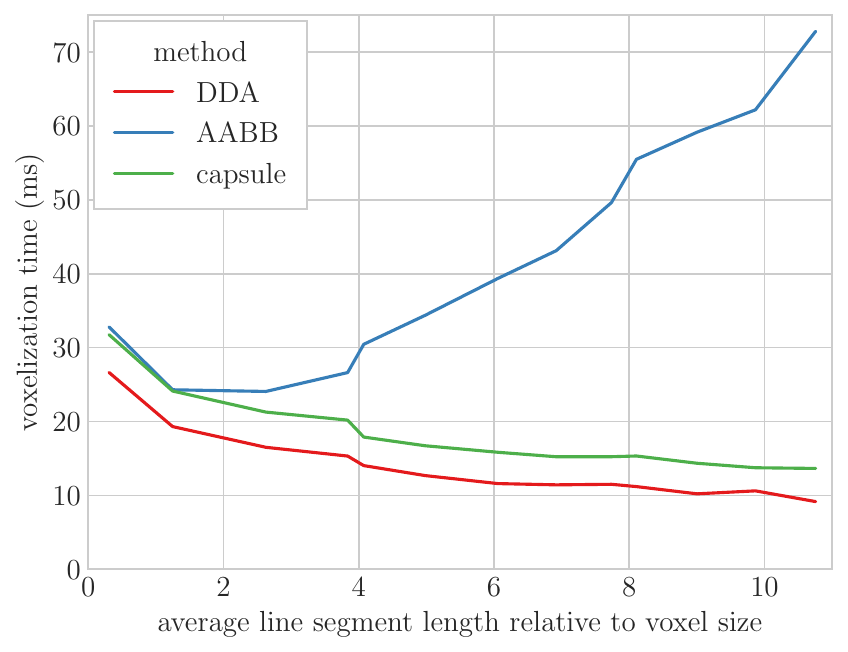}
    \caption{Voxelization performance as a function of segment length.}
    \label{fig:results:voxelization:performance:line-length}
    \vspace{-\baselineskip}
\end{figure}

To further investigate the performance of the three voxelization strategies, we evaluated the effect of line segment length on performance, as shown in \cref{fig:results:voxelization:performance:line-length}.
We computed various levels of detail of the high-resolution \textit{Whole Brain 1M} dataset by connecting every \(n\)th vertex to produce level of detail \(n\).
The resulting line sets span roughly the same number of voxels but have increasing average line segment lengths.
For all tests, we measured voxelization performance using a resolution of \(256^3\) and a line radius of \(0.2\) times the voxel size.
\cref{fig:results:voxelization:performance:line-length} shows the advantage of \textit{capsule} voxelization algorithm over the \textit{AABB} approach for longer line segments.
\textit{Capsule} voxelization adds a fixed overhead compared to \textit{DDA} voxelization.
These findings corroborate the theoretical efficiencies in \cref{method:voxelization}.

% Experiment 2: Line Length x Performance (Line vs Box vs Capsule)
\begin{figure*}
    \centering
    \begin{subfigure}[t]{0.33\linewidth}
        \centering
        \includegraphics[width=\linewidth]{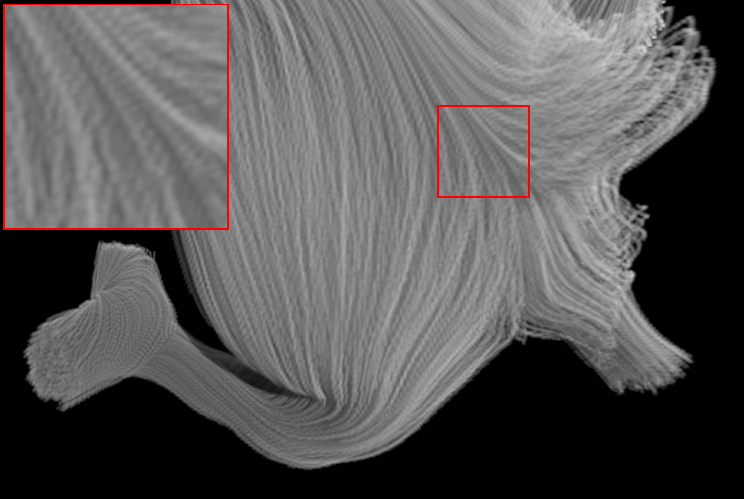}
        \caption{DDA occupancy}
        \label{fig:results:voxelization:clip:line}
    \end{subfigure}%
    \begin{subfigure}[t]{0.33\linewidth}
        \centering
        \includegraphics[width=\linewidth]{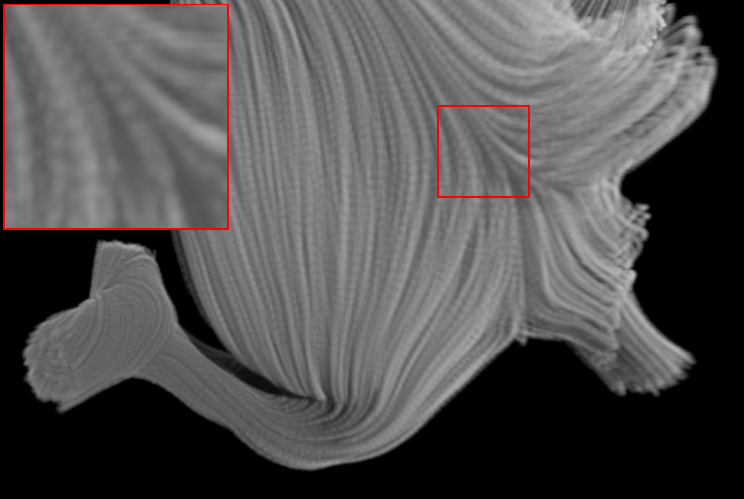}
        \caption{SDF occupancy without clipping planes}
        \label{fig:results:voxelization:clip:without}
    \end{subfigure}%
    \begin{subfigure}[t]{0.33\linewidth}
        \centering
        \includegraphics[width=\linewidth]{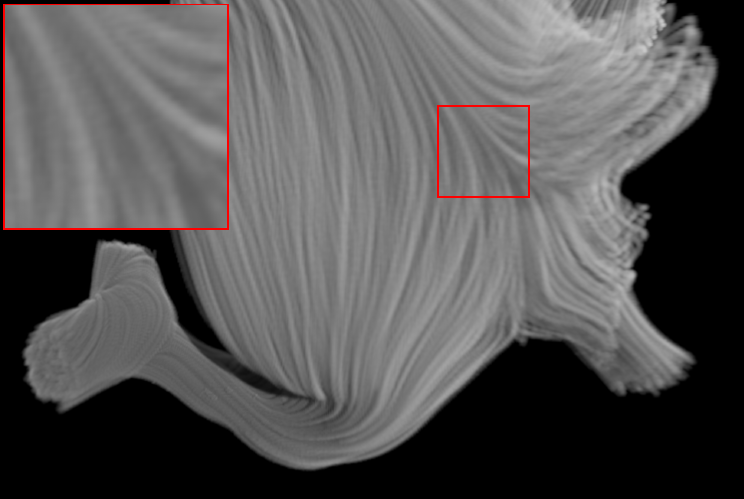}
        \caption{SDF occupancy with clipping planes}
        \label{fig:results:voxelization:clip:with}
    \end{subfigure}
    \caption{
        The effect of clipping planes on voxelization quality.
        Without clipping planes, spherical artifacts appear at interior vertices.
        Using the clipped capsule SDF removes these artifacts and significantly reduces aliasing compared to DDA voxelization.}
    \label{fig:results:voxelization:clip}
\end{figure*}

\begin{figure*}
    \centering
    \begin{subfigure}[t]{0.33\linewidth}
        \centering
        \includegraphics[width=\linewidth]{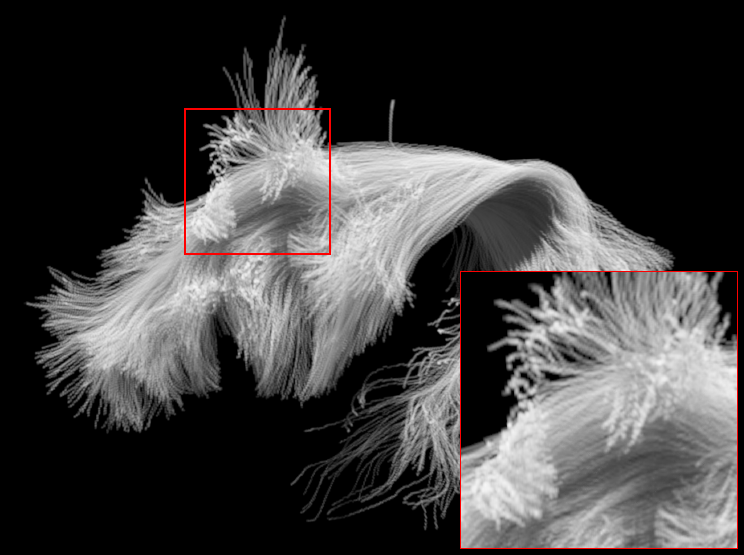}
        \caption{DDA occupancy}
        \label{fig:results:voxelization:pwaa:line}
    \end{subfigure}%
    \begin{subfigure}[t]{0.33\linewidth}
        \centering
        \includegraphics[width=\linewidth]{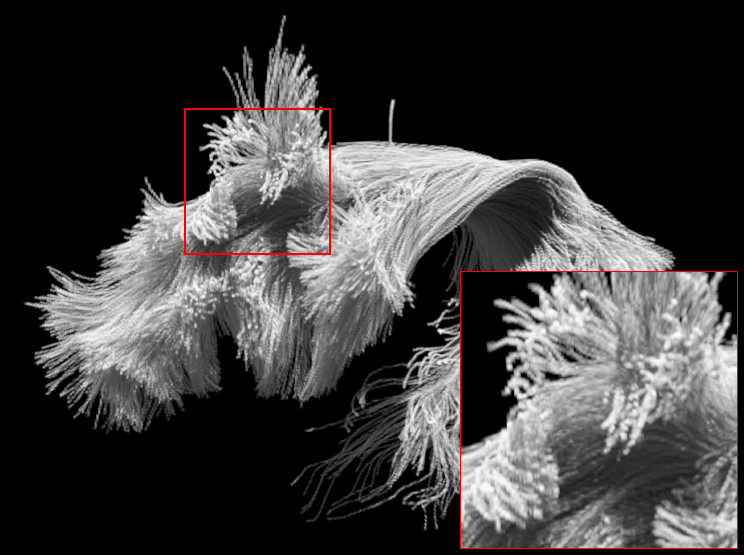}
        \caption{SDF occupancy without PWAA}
        \label{fig:results:voxelization:pwaa:without}
    \end{subfigure}%
    \begin{subfigure}[t]{0.33\linewidth}
        \centering
        \includegraphics[width=\linewidth]{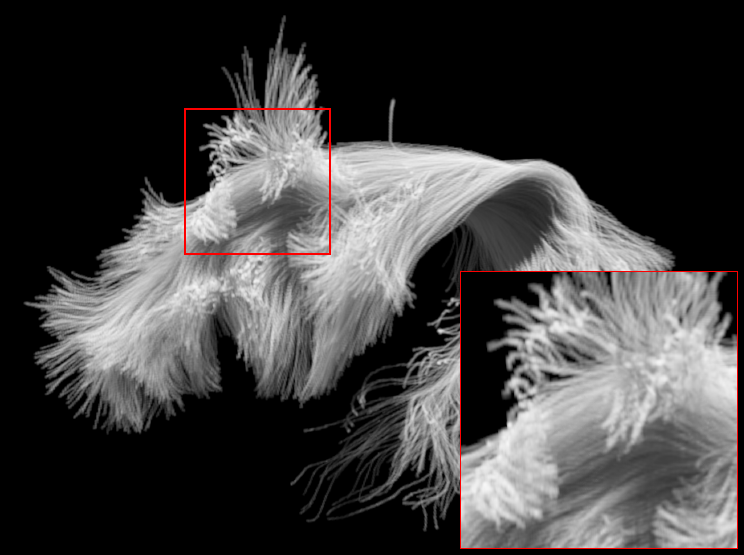}
        \caption{SDF occupancy with PWAA}
        \label{fig:results:voxelization:pwaa:with}
    \end{subfigure}
    \caption{
        The effect of Phone-Wire Anti-Aliasing on voxelization quality.
        Without Phone-Wire Anti-Aliasing, SDF-based voxelization of thin lines introduces aliasing and an overestimation of total occupancy.
        Applying Phone-Wire Anti-Aliasing reduces these artifacts.}
    \label{fig:results:voxelization:pwaa}
\end{figure*}

% Qualitative Analysis
Now that we have investigated the performance of our voxelization strategy, we examine the qualitative aspects of voxelization.
Since \textit{DDA} only visits voxels that are directly intersected by an infinitely thin line segment, occupancy estimation is restricted to those voxels, leading to aliasing for larger line radii.
Conservative voxelization algorithms visit all voxels intersected by the capsule primitive, allowing for more accurate occupancy estimation.
We evaluate how clipping planes and Phone-Wire Anti-Aliasing (PWAA) improve SDF-based occupancy estimation compared to \textit{DDA}.

% Voxelization Quality: Clipping
\cref{fig:results:voxelization:clip} compares DDA-based occupancy and SDF-based occupancy both with and without clipping planes.
The DDA-based occupancy, shown in \cref{fig:results:voxelization:clip:line}, clearly exhibits aliasing for larger line radii.
SDF-based occupancy without clipping planes, shown in \cref{fig:results:voxelization:clip:without}, reduces aliasing but introduces spherical artifacts at every interior vertex.
These artifacts cause an overestimation of occupancy, especially when voxelizing short segments with large radii.
Our clipped capsule signed distance function, shown in \cref{fig:results:voxelization:clip:with}, removes the artifacts and reduces aliasing compared to \textit{DDA}.

% Voxelization Quality: Phone Wire Anti-Aliasing
\cref{fig:results:voxelization:pwaa} compares DDA-based occupancy with SDF-based occupancy with and without Phone-Wire Anti-Aliasing (PWAA).
Shown in \cref{fig:results:voxelization:pwaa:line}, DDA-based occupancy estimation produces minimal aliasing for thin line segments.
However, SDF-based occupancy without PWAA, shown in \cref{fig:results:voxelization:pwaa:without}, introduces severe aliasing.
PWAA, shown in \cref{fig:results:voxelization:pwaa:with}, reduces aliasing and matches or slightly improves upon the results of DDA-based occupancy.

\begin{figure}
    \centering
    \begin{subfigure}[t]{\linewidth}
        \centering
        \includegraphics[width=0.9\linewidth]{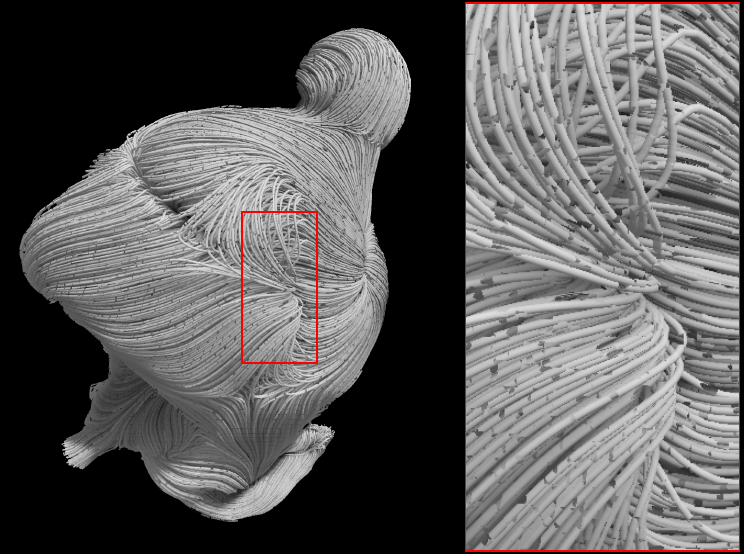}
        \caption{Non-Conservative Voxelization}
    \end{subfigure}
    \begin{subfigure}[t]{\linewidth}
        \centering        
        \includegraphics[width=0.9\linewidth]{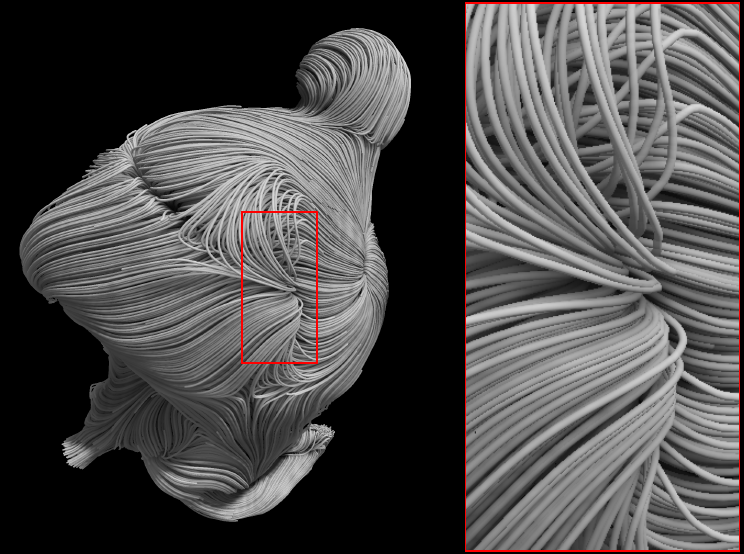}
        \caption{Conservative Voxelization}
    \end{subfigure}
    \caption{
    Comparison of ray tracing using conservative versus non-conservative voxelization for the Aneurysm dataset. Cube-like artifacts are visible when employing non-conservative voxelization. Conservative voxelization correctly renders the capsule primitives.}
    \label{fig:results:voxelization:raytracing-artifacts}
\end{figure}

% Voxelization and Ray-Tracing Artifacts
\cref{fig:results:voxelization:raytracing-artifacts} illustrates missing-segment artifacts during ray tracing, which occur when a ray passes through a voxel that does not contain the intersecting primitive.
These artifacts appear as cubical cutouts and become more apparent with larger line radii relative to voxel size.
Conservative voxelization completely eliminates these artifacts.

% Conclusion
To conclude, these results show that the choice of voxelization algorithm depends on the scenario.
\textit{DDA} voxelization offers the best performance and shows minimal aliasing for thin line segments but is not conservative.
Conservative voxelization should be used for ray tracing and for thicker line segments.
As expected, \textit{capsule} voxelization scales linearly with segment length, compared to the worst-case cubic time complexity of \textit{AABB} voxelization.

\subsection{Per-Voxel Fragment Lists} \label{results:abuffer}

We capture per-voxel fragment lists using a two-pass voxelization scheme: \textit{voxelize-scan-voxelize} (VSV).
To improve performance for (semi-)opaque line sets, we introduced a culling step to accelerate the secondary voxelization pass: \textit{voxelize-cull-scan-voxelize} (VCSV).
We compare these two approaches to another common method: \textit{voxelize-sort-scan} (VSS) \cite{kanzler2018, mcgraw2020}, using OneSweep \cite{adinets2022onesweep} as a fast sorting implementation.
In \cref{tab:results:abuffer:performance} we compare these strategies to compute both the occupancy pyramid and the per-voxel fragment lists, i.e., all per-frame preprocessing required for dynamic ray tracing and ambient occlusion.

\begin{table}
    \centering
    \begin{tabular}{l|ccc}
    \textbf{Dataset} & \textbf{VSS} & \textbf{VSV (ours)} & \textbf{VCSV (ours)} \\
    \hline
    Bundles Small       & 14    & 5     & 4    \\
    Aneurysm            & 21    & 6     & 4    \\
    Bundles Large       & 75    & 18    & 12   \\
    Brain 200k          & 67    & 20    & 12   \\
    Turbulence          & 120   & 46    & 21   \\
    Brain 1M            & 1015  & 104   & 50   \\
    \end{tabular}
    \caption{
        Performance in milliseconds for generating both the \textit{occupancy pyramid} and the \textit{per-voxel fragment lists} at a resolution of \(128^3\) for opaque geometry, using: \textit{voxelize-sort-scan} (VSS), \textit{voxelize-scan-voxelize} (VSV) and \textit{voxelize-cull-scan-voxelize} (VCSV).}
    \label{tab:results:abuffer:performance}
\end{table}

As expected, \textit{voxelize-scan-voxelize} (VSV) outperforms \textit{voxelize-sort-scan} (VSS), while producing the same per-voxel fragment lists and occupancy pyramid.
The threefold speedup aligns with the theoretical advantage of VSV.
VSV requires only two passes over all fragments and one scan over all voxels, whereas VSS needs seven passes over all fragments (one write, five for sorting, one scan).
While Morton-order sorting generates a cache-efficient acceleration structure, similar efficiency can be achieved by the choice of scan pass in the \textit{voxelize-scan-voxelize} strategy.

Another advantage of the \textit{voxelize-scan-voxelize} (VSV) strategy is the potential of accelerating the secondary voxelization pass.
The occupancy pyramid generated in the primary voxelization pass enables view-dependent culling of the secondary voxelization pass, resulting in the \textit{voxelize-cull-scan-voxelize} (VCSV) strategy.
When rendering opaque geometry, VCSV improves performance up to two times compared with VSV.
The exact gain depends on line set density and camera angle.
Altogether, our method cuts down per-frame preprocessing time by a factor of three when rendering transparent data (VSV) or a factor of six for opaque data.

\subsection{Rendering Pipeline} \label{results:rendering}

Next, we compare our dynamic rendering pipeline to the methods of Kanzler et al. \cite{kanzler2018} and Gro{\ss} and Gumhold \cite{gross2020}, who both render line sets with ambient occlusion and transparency.
We test the dynamic aspect of our method and these state-of-the-art methods by including all per-frame preprocessing costs in all performance measurements.
By building all acceleration structures from scratch every frame, all rendering pipelines can handle arbitrary dynamic behavior.
Since these state-of-the-art methods rely on fast sorting, we employ the OneSweep radix-sort algorithm \cite{adinets2022onesweep}.

Gro{\ss} and Gumhold's rasterization method \cite{gross2020} handles dynamic data, except for the construction of the occupancy pyramid.
We therefore generate the occupancy pyramid every frame using fast DDA voxelization, making their method fully dynamic.
Otherwise, we keep all aspects the same as in the original paper.

We use the GPU quantization method of Kanzler et al. \cite{kanzler2018} to construct the occupancy pyramid and per-voxel fragment lists every frame, making their method fully dynamic.
We use a quantization resolution of \(64^2\), which makes each segment fit in 32 bits for optimal performance.
While the authors included a level-of-detail system, we omit it in our experiments, because such a system could be implemented for both Gro{\ss} and Gumhold and our method.
Because it reduces performance, we exclude the authors’ voxel neighborhood sampling in our performance and quality evaluation.
As a performance reference, we compare these methods to an opaque rasterization \textit{Baseline}.
This baseline implements as Gro{\ss} and Gumhold's method \cite{gross2020} without sorting and culling, using DDA to compute the occupancy pyramid.

All experiments target a screen resolution of \(1920 \times 1080\).
We measure the time it takes to complete the entire dynamic rendering pipeline, assuming that all acceleration structures are recomputed every frame for each method.
We chose a voxel resolution of \(128^3\), which provides optimal ray-tracing performance according to our experiments and does not affect line rendering quality for our method.
We use a line radius of 0.2 times the voxel size.

\begin{figure*}
    \centering
    \begin{subfigure}[t]{0.32\linewidth}
        \centering
        \includegraphics[width=\linewidth]{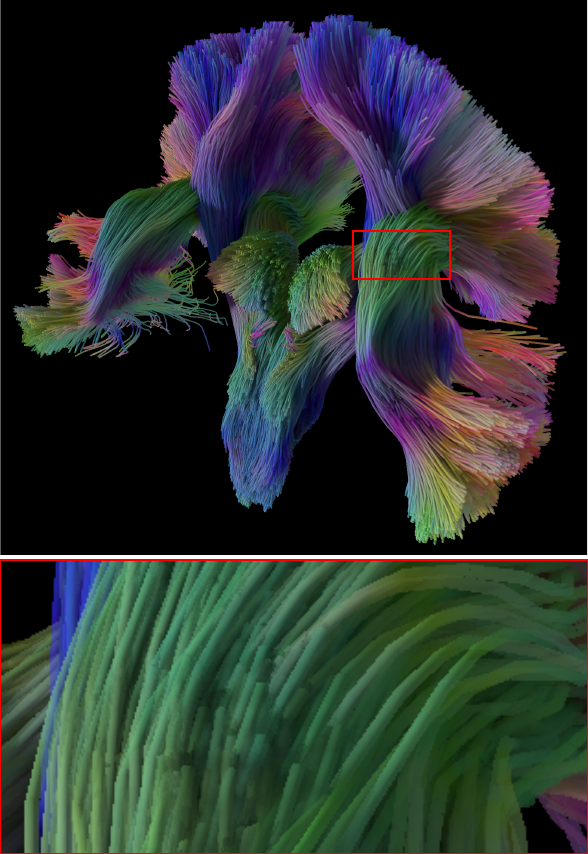}
        \caption{Gro{\ss} and Gumhold (2021)}
    \end{subfigure}
    \begin{subfigure}[t]{0.32\linewidth}
        \centering
        \includegraphics[width=\linewidth]{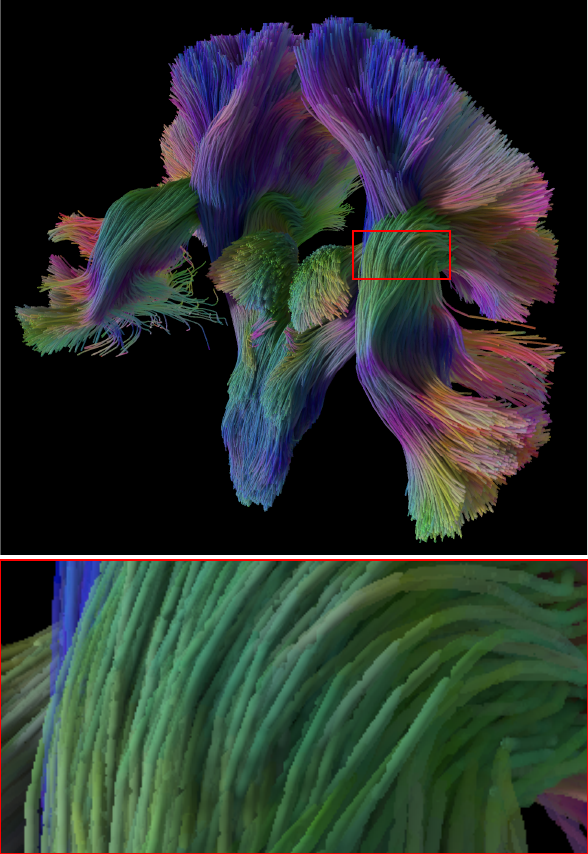}
        \caption{Kanzler et al. (2019)}
    \end{subfigure}
    \begin{subfigure}[t]{0.32\linewidth}
        \centering
        \includegraphics[width=\linewidth]{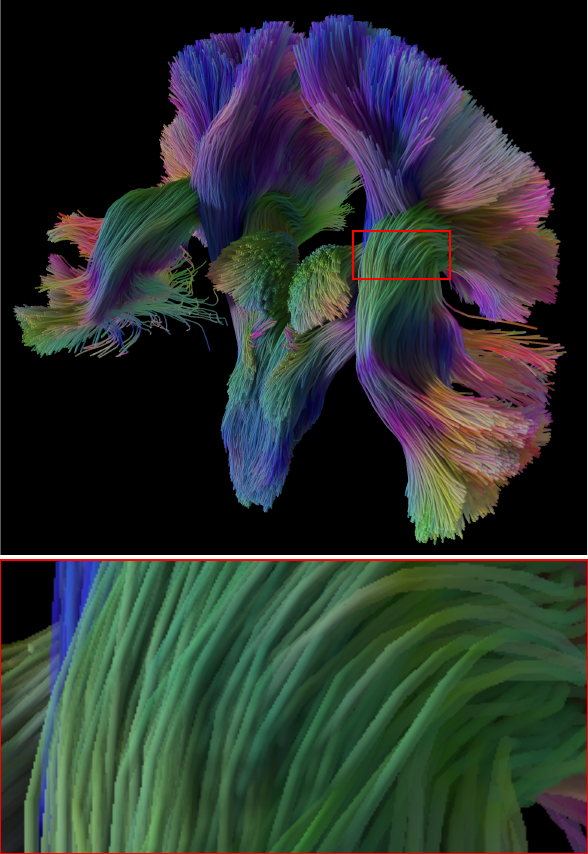}
        \caption{Ours}
    \end{subfigure}
    \caption{
        Qualitative comparison between rendering methods.
        For the Bundles Small dataset, all methods produce similar renderings. However, Gro{\ss} and Gumhold's method suffers from sorting artifacts and Kanzler et al.'s method exhibits missing-segment artifacts. Our voxel-based ray-tracing method resolves both types of artifacts.}
    \label{fig:results:rendering:quality}
\end{figure*}

First, we evaluate rendering quality using semi-opaque renderings of the \textit{Bundles Small} line set, shown in \cref{fig:results:rendering:quality}, which presents a challenge for the state of the art.
All methods achieve similar overall renderings, but closer inspection reveals qualitative differences.
While their method handles shorter line segments well, Gro{\ss} and Gumhold's K-Buffer-based order-correcting algorithm fails to properly sort the longer line segments in this line set.
Being non-conservative, Kanzler et al.'s method shows missing-segment artifacts, which reduces the rendering quality for (semi-)opaque lines.
These artifacts are less noticeable when rendering fully transparent geometry.
Our voxel-based ray tracing guarantees correct depth ordering and eliminates missing-segment artifacts.

% Rendering Quality

% Opaque Rendering Performance
As shown in \cref{tab:results:rendering:performance:opaque}, our method achieves the highest render performance across all dynamic opaque line sets.
Our method is closely followed by that of Kanzler et al.; however, their method exhibits quantization and missing-segment artifacts.
Our performance gains are attributable to our proposed acceleration structure construction method, which is especially effective when rendering opaque line sets.
Compared to the rasterization techniques (\textit{Baseline} and \textit{Gro{\ss}}), the ray-tracing techniques (\textit{Kanzler} and \textit{Ours}) have the advantage of early ray termination and show superior performance on larger line sets.
As expected, all methods outperform \textit{Baseline} rasterization.

\begin{table}
    \centering
    \begin{tabular}{l|ccccc}
        \textbf{Dataset} & \textbf{Baseline} & \textbf{Gro{\ss}}  & \textbf{Kanzler} & \textbf{Ours} \\
        \hline
        Bundles Small   &  13 &  12 &  13 & \textbf{10} \\
        Aneurysm        &  25 &  20 &  13 & \textbf{9} \\ 
        Bundles Large   &  83 &  44 &  39 & \textbf{22} \\
        Brain 200k      & 138 &  81 &  30 & \textbf{28} \\ 
        Turbulence      & 209 & 127 &  47 & \textbf{41} \\ 
        Brain 1M        & 822 & 417 & 119 & \textbf{117} \\
        \end{tabular}
    \caption{
        Opaque rendering performance, measured in milliseconds, including all per-frame preprocessing costs.}
    \label{tab:results:rendering:performance:opaque}
\end{table}

% Transparent Rendering Performance
\begin{table}
    \centering
    \begin{tabular}{l|ccc}
        \textbf{Dataset} & \textbf{Gro{\ss}} & \textbf{Kanzler} & \textbf{Ours} \\
        \hline
        Bundles Small   & \textbf{32}   & 35 &  54 \\
        Aneurysm        & 39            & \textbf{30}   &  55 \\ 
        Bundles Large   & 81            & \textbf{67}   &  89 \\
        Brain 200k      & 127           & \textbf{55}   & 117 \\ 
        Turbulence      & 175           & \textbf{97}   & 199 \\ 
        Brain 1M        & 596           & \textbf{158}  & 276 \\
        \end{tabular}
    \caption{
        Transparent rendering performance (opacity 0.1), measured in milliseconds, including all per-frame preprocessing costs.}
    \label{tab:results:rendering:performance:transparent}
\end{table}

\cref{tab:results:rendering:performance:transparent} shows rendering performance for transparent geometry using an opacity of 0.1.
When rendering transparent geometry, Kanzler et al. show the best performance for most datasets.
While our method performs well on opaque line sets, performance is significantly reduced when rendering full transparency.
This performance reduction is expected, since conservative voxelization increases the number of fragments and ray-capsule intersection tests, which dominate the rendering time for fully transparent line sets.
Kanzler et al.'s quantized method increases performance by limiting the number of ray-capsule intersections compared to more accurate methods.

% Rendering Performance versus Opacity
\begin{figure}
    \centering
    \includegraphics[width=0.9\linewidth]{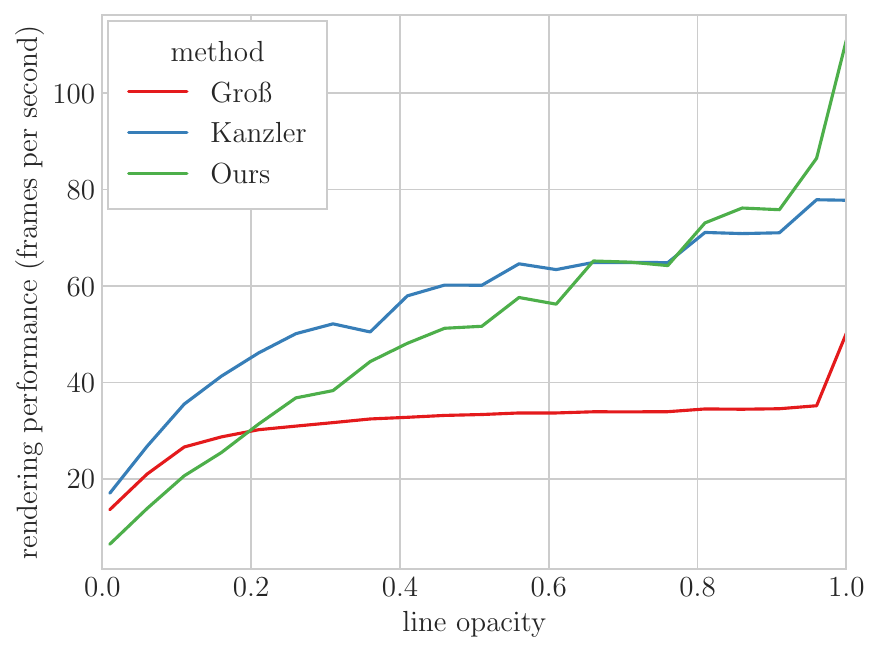}
    \caption{
        Rendering performance at varying line opacities for the Aneurysm line set, in frames per second, including all preprocessing.}
    \label{fig:results:rendering:performance}
\end{figure}

To investigate the impact of line opacity on dynamic rendering performance, we evaluated the rendering performance of the \textit{Aneurysm} line set at various opacities, as shown in \cref{fig:results:rendering:performance}.
This figure clearly shows the advantage of early ray-termination over the hierarchical culling strategy of Gro{\ss} and Gumhold.
Given that our method renders correct transparency without quantization and artifacts, it performs favorably compared to the method of Kanzler et al., even surpassing their method for very high line opacities.

Finally, a performance breakdown of our rendering pipeline is shown in \cref{appendix:breakdown}, reporting the performance of each step for opaque geometry.
As in our other experiments, we chose a grid resolution of \(128^3\), a screen resolution of \(1920 \times 1080\), and a line radius of \(0.2\) times the voxel size.
Culling only adds a small, fixed overhead for our chosen grid resolution, since it is performed in voxel space.
The per-voxel fragment list generation step remains a major contributor in the per-frame preprocessing time, even after our optimizations.
We also perform shading in voxel space, which adds a small overhead depending on the number of visible voxels.
Precomputation still accounts for most of the total rendering time for opaque geometry.
When rendering static line sets, only the ray-tracing step needs to be computed every frame, which results in about a twofold speedup compared to dynamic line sets.

%-------------------------------------------------------------------------
\section{Conclusion and Future Work}

% Recap
We have presented a real-time rendering pipeline for voxel-based ray tracing of dynamic line sets.
Our method improves rendering quality by introducing a conservative capsule voxelization method that supports ray tracing.
We improve rendering performance by using a view-dependent culling approach to efficiently compute per-voxel fragment lists.
By using our voxel ray tracing technique, we render (semi-)opaque line sets with ambient occlusion and ground truth transparency in real time.

% Limitations
A limitation of our approach is the low rendering performance when rendering lines at low opacity values.
This may be improved by adding level-of-detail rendering or by exploiting frame coherence.
In the future, we will investigate how interactive diffusion MRI tractography segmentation can benefit from our rendering approach.

\section*{Acknowledgments}
This publication is part of the project \textit{VIBRANT} with file number \textit{OCENW.M.22.352} of the research programme \textit{Open Competition Domain Science – M}, which is (partly) financed by the Dutch Research Council (NWO).
\begin{appendix}

% \appendix
\crefalias{section}{appendix}

\begin{figure*}[t]

    \section{Line Set Renderings} \label{appendix:datasets}
    
    \vspace{\baselineskip}

    \centering
    \begin{subfigure}[t]{0.33\linewidth}
        \centering
        \includegraphics[width=\linewidth]{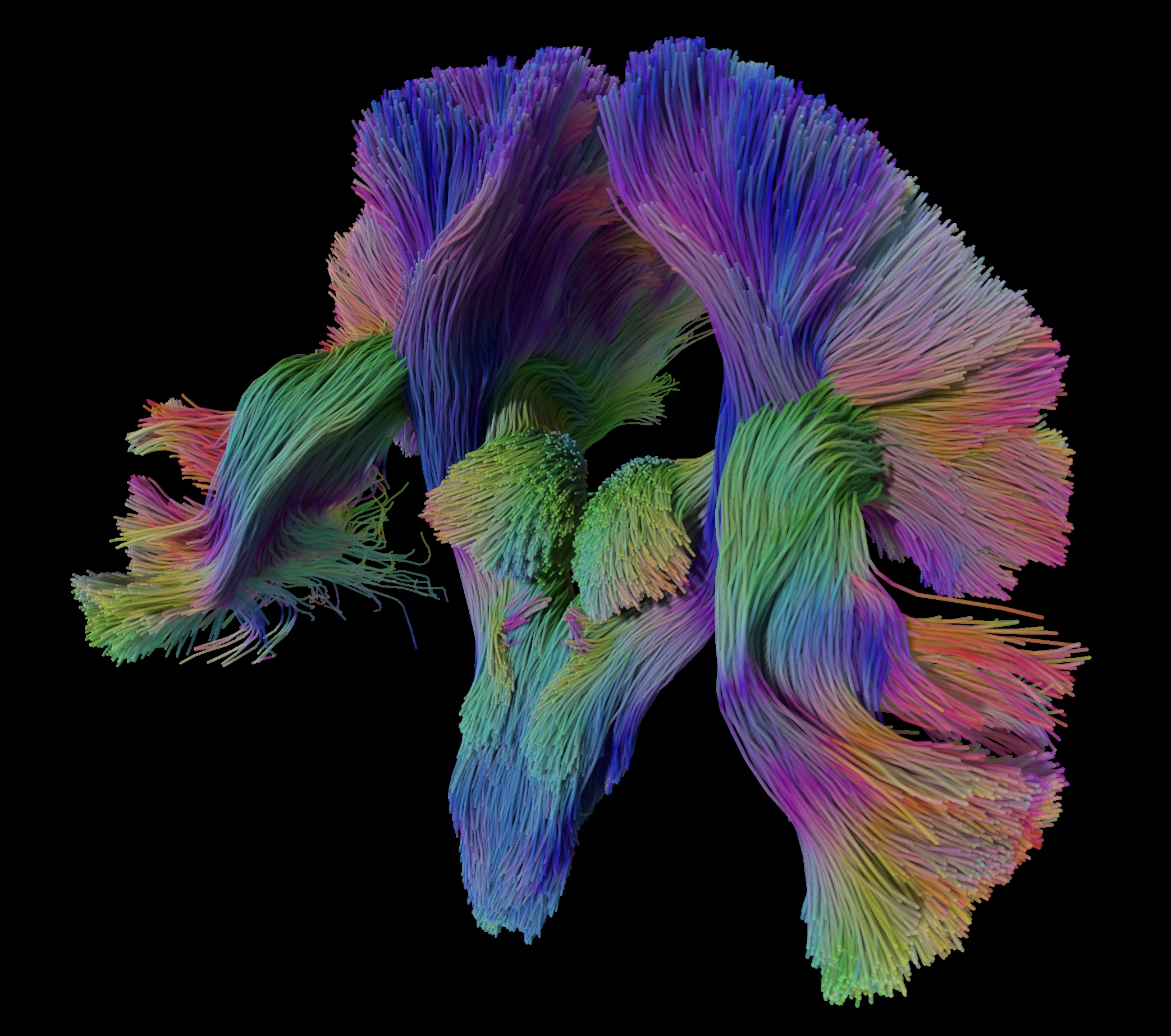}
        \caption{Bundles Small}
        \label{fig:appendix:bundles-small}
    \end{subfigure}%
    \begin{subfigure}[t]{0.33\linewidth}
        \centering
        \includegraphics[width=\linewidth]{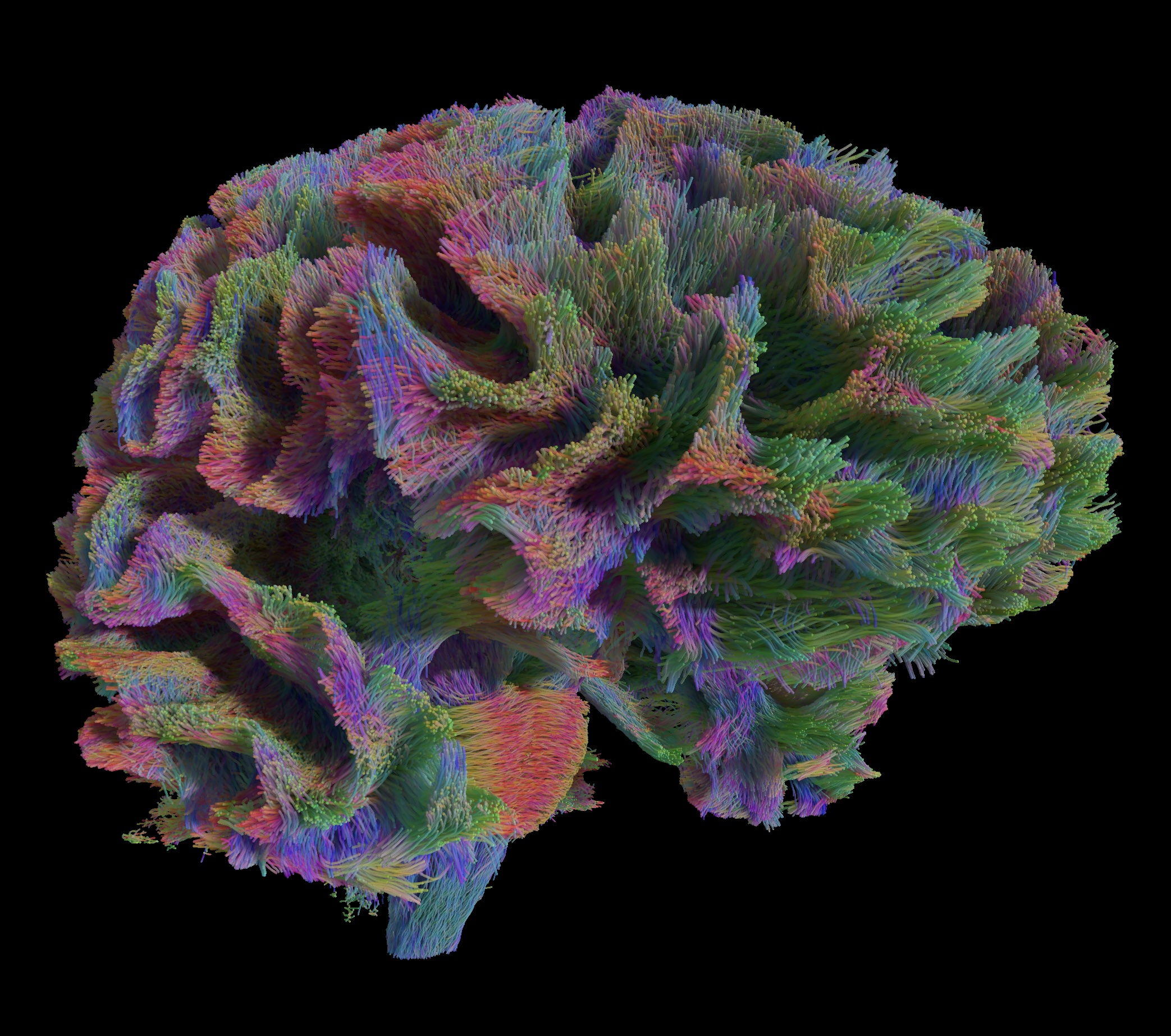}
        \caption{Brain 200k}
        \label{fig:appendix:brain-200k}
    \end{subfigure}%
    \begin{subfigure}[t]{0.33\linewidth}
        \centering
        \includegraphics[width=\linewidth]{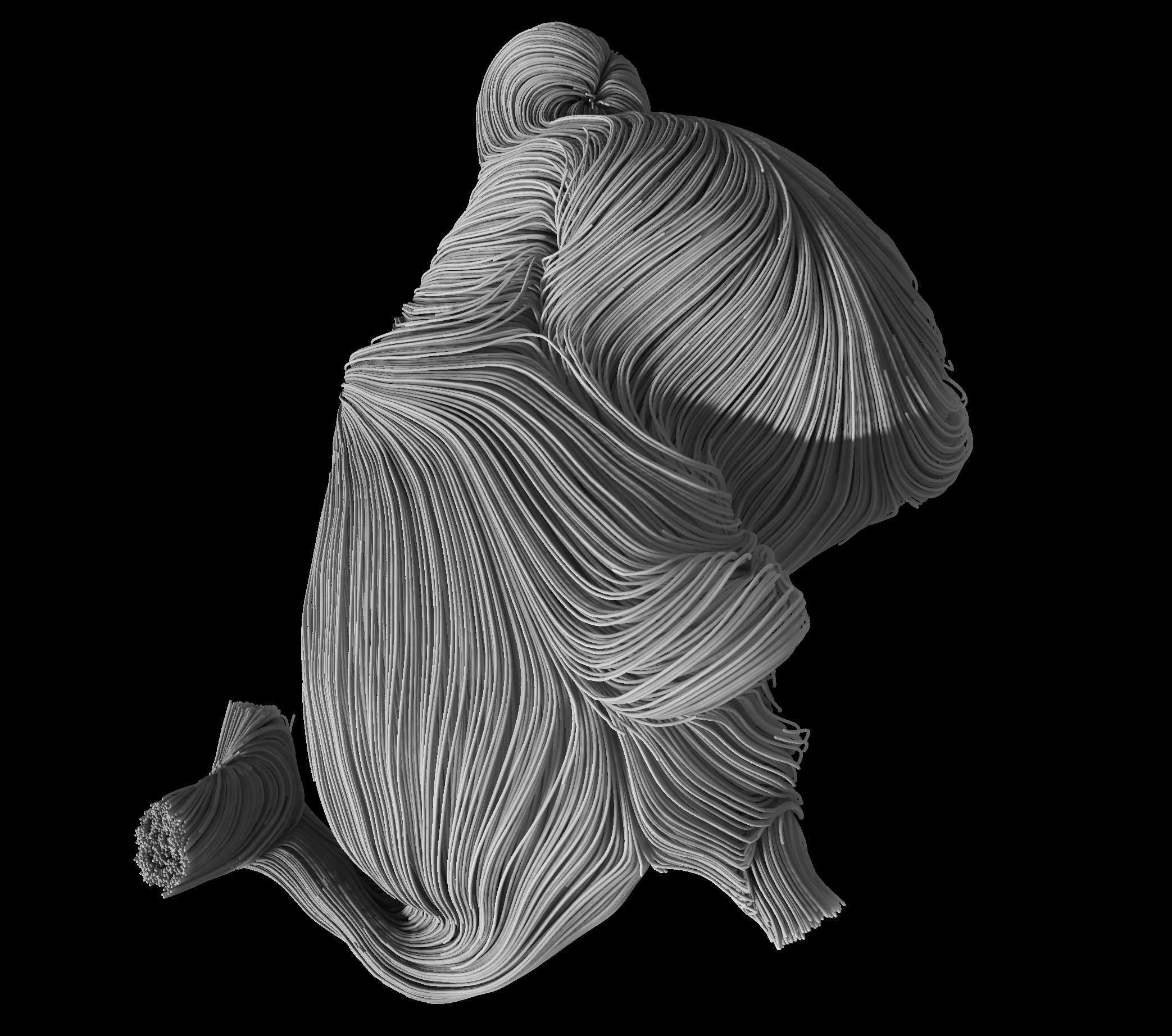}
        \caption{Aneurysm\vspace{\baselineskip}}
        \label{fig:appendix:aneurysm}
    \end{subfigure}
    \begin{subfigure}[t]{0.33\linewidth}
        \centering
        \includegraphics[width=\linewidth]{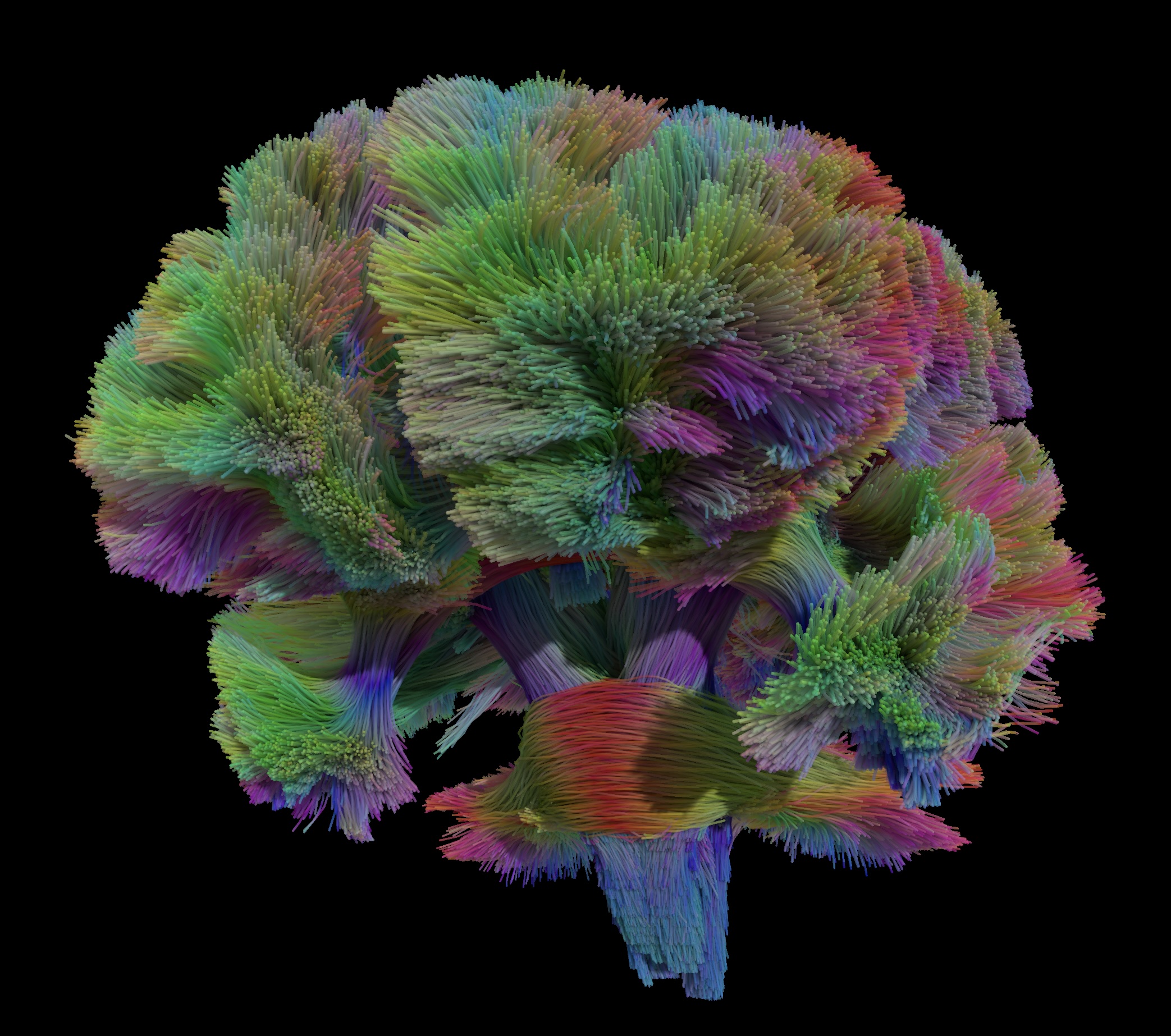}
        \caption{Bundles Large}
        \label{fig:appendix:bundles-big}
    \end{subfigure}%
    \begin{subfigure}[t]{0.33\linewidth}
        \centering
        \includegraphics[width=\linewidth]{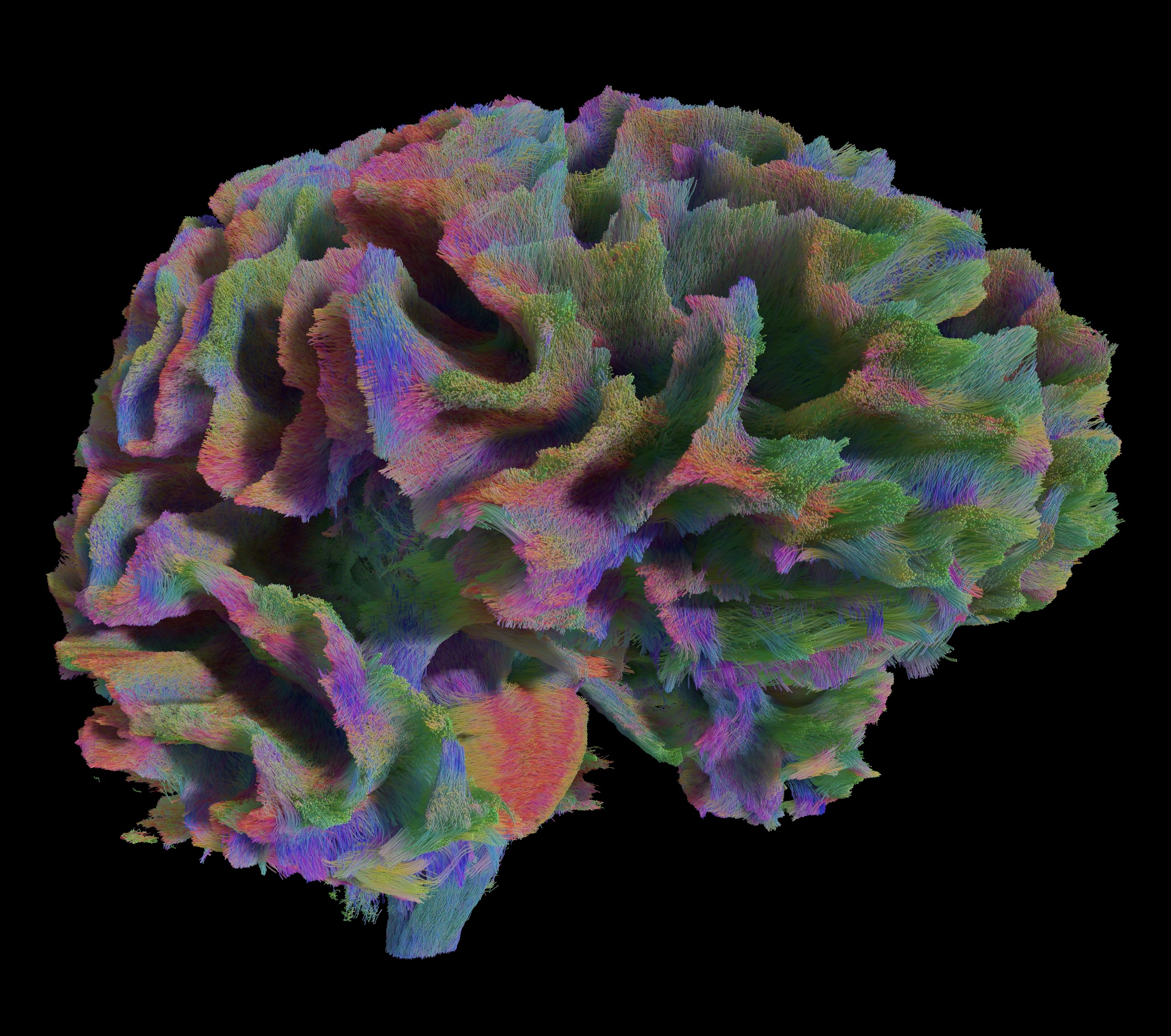}
        \caption{Brain 1M}
        \label{fig:appendix:brain-1M}
    \end{subfigure}%
    \begin{subfigure}[t]{0.33\linewidth}
        \centering
        \includegraphics[width=\linewidth]{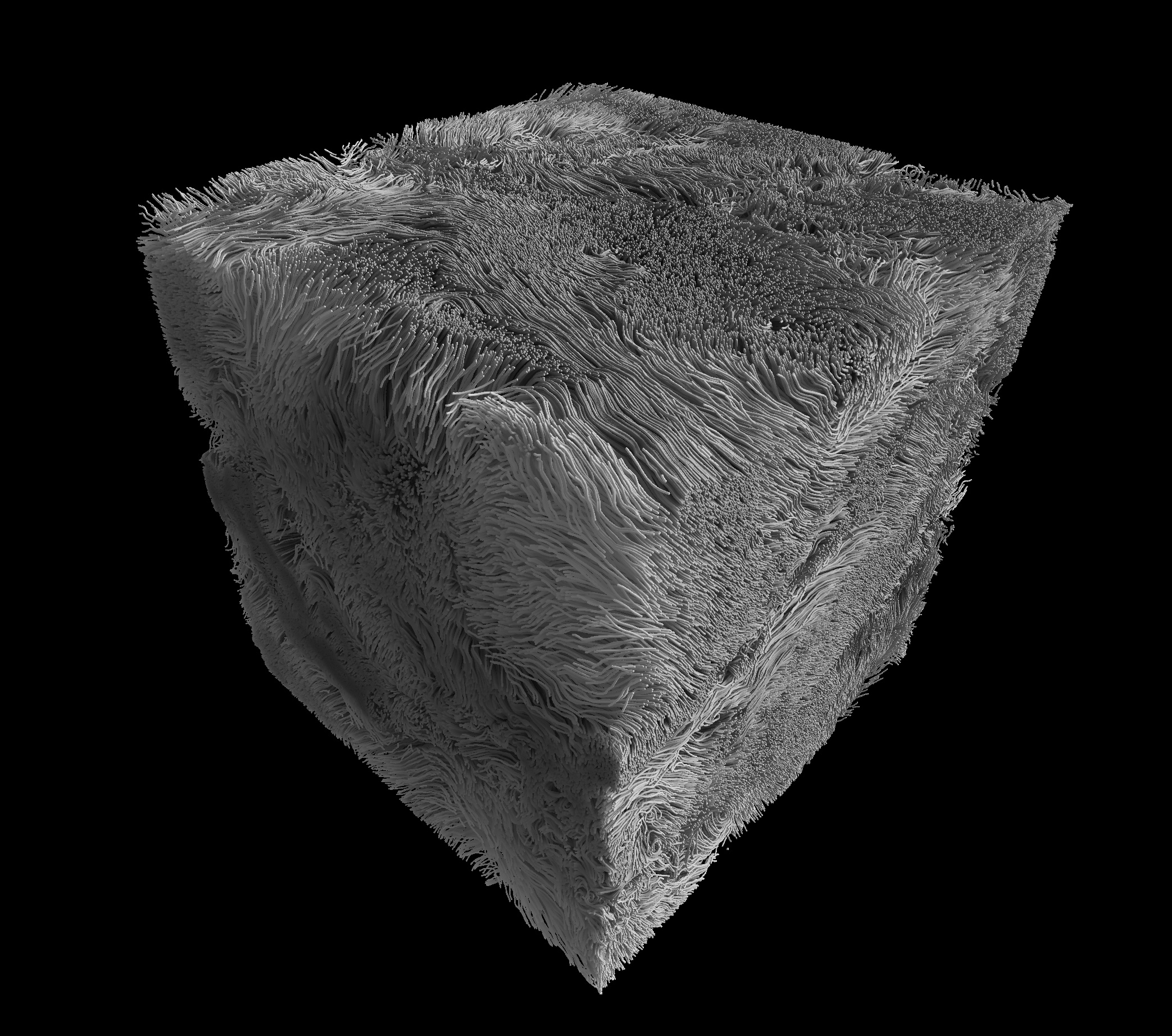}
        \caption{Turbulence\vspace{\baselineskip}}
        \label{fig:appendix:turbulence}
    \end{subfigure}
\end{figure*}

\begin{table*}[h!]
    \section{Performance Breakdown} \label{appendix:breakdown}
    \vspace{\baselineskip}

    \centering
    \begin{tabular}{c|cccc|cc}
    \textbf{Dataset} & \textbf{Voxelization} & \textbf{Culling} & \textbf{Per-Voxel Fragment List} & \textbf{Shading} & \textbf{Total Precompute} & \textbf{Ray Tracing} \\
    \hline
       Bundles Small    & 3.6   & 1.3   & 2.7   & 1.3   & 8.9   & 5.1   \\
       Aneurysm         & 2.7   & 1.3   & 2.7   & 2.4   & 9.1   & 4.9   \\
       Bundles Large    & 6.1   & 1.3   & 6.1   & 3.4   & 16.9  & 7.7   \\
       Brain 200k       & 5.2   & 1.3   & 7.7   & 1.3   & 15.5  & 10.3  \\
       Turbulence       & 8.2   & 1.3   & 12.7  & 2.9   & 25.1  & 11.6  \\
       Brain 1M         & 20.0  & 1.3   & 30.2  & 2.9   & 54.4  & 32.6  \\
    \end{tabular}
    \caption{
        Performance, in milliseconds, for each stage of our rendering pipeline, corresponding with \cref{method:voxelization,method:culling,method:abuffer,method:shading,method:rendering}, for dynamic opaque line sets. When rendering static line sets, only the ray-tracing step needs to be computed each frame.}
\end{table*}

\end{appendix}

\newpage
\printbibliography

\end{document}